\documentclass[traditabstract]{aa}  

\usepackage{graphicx}
\usepackage{geometry}
\usepackage{pdflscape}
\usepackage[usenames,dvipsnames]{color} 

\usepackage{txfonts}
\usepackage{natbib}
\usepackage{longtable}
\usepackage{aalongtable}
\usepackage{lscape}
\usepackage{multicol}
\usepackage{subfig}
\usepackage{caption} 
\usepackage{placeins}

\bibpunct{(}{)}{;}{a}{}{,} 
\def\plm{$\pm$}

\newcommand{\teff}{$T_\mathrm{eff}$}
\newcommand{\logg}{\ensuremath{\log g}}

\newcommand{\vt}{v$_{\rm turb}$}
\newcommand{\kms}{\,km\,s$^{-1}$}         

\usepackage{color}

\begin{document}

   \title{The chemical evolution of the dwarf spheroidal galaxy Sextans\thanks{Based on the ESO Program 171.B-0588(A)}\thanks{Tables 2-6 and 9-13 are available in electronic form at the CDS via anonymous ftp to cdsarc.u-strasbg.fr}}
   

   \author{R. Theler\inst{1} \and P. Jablonka\inst{1,2}  \and R. Lucchesi\inst{1}  \and C. Lardo\inst{1} \and P. North\inst{1} \and M. Irwin \inst{3}  \and Battaglia G. \inst{4,5}  \and V. Hill \inst{6} \and E. Tolstoy\inst{7}  \and K. Venn\inst{8} \and  A. Helmi\inst{7} \and  A. Kaufer\inst{9} \and F. Primas\inst{10} \and  Shetrone M.\inst{11} }
     
     \institute{Physics Institute, Laboratory of Astrophysics, Ecole Polytechnique F\'ed\'erale de Lausanne (EPFL), 1290 Sauverny, Switzerland \email{pascale.jablonka@epfl.ch}
       \and GEPI, Observatoire de Paris, Universit\'e PSL, CNRS, Place Jules Janssen, F-92190 Meudon, France
       \and Institute of Astronomy, University of Cambridge, Madingley Road, Cambridge CB3 0HA, U.K.
       \and Instituto de Astrof\'isica de Canarias (IAC), Calle Via L\'actea, s/n, 38205, San Crist\'obal de la Laguna, Tenerife, Spain
       \and Departamento de Astrof\'isica, Universidad de La Laguna, 38206, San Crist\'obal de la Laguna, Tenerife, Spain
       \and Laboratoire Lagrange, Universit\'e de Nice Sophia-Antipolis, Observatoire de la C\^ote d'Azur, France
       \and Kapteyn Astronomical Institute, University of Groningen, Landleven 12, NL-9747AD Groningen, the Netherlands
       \and Department of Physics and Astronomy, University of Victoria, PO Box 3055, STN CSC, Victoria BC V8W 3P6, Canada
       \and European Southern Observatory, Alonso de Cordova 3107, Vitacura, Casilla 19001, Santiago, Chile
       \and European Southern Observatory, Schwarzschild-Str. 2, 85748 Garching, Germany
       \and McDonald Observatory, University of Texas at Austin, Fort David, TX, USA
     }

   \date{Received XX XX, XXXX; accepted XX XX, XXXX}

   \abstract{We present our analysis of the FLAMES dataset targeting
     the central 25\arcmin\ region of the Sextans dwarf spheroidal  galaxy (dSph). This dataset
     is the third major part of the high-resolution spectroscopic
     section of the ESO large program 171.B-0588(A) obtained by the
     Dwarf galaxy Abundances and Radial-velocities Team (DART). Our
     sample is composed of red giant branch stars down to V$\sim$20.5
     mag, the level of the horizontal branch in Sextans, and allows users to
     address questions related to both stellar nucleosynthesis and
     galaxy evolution.

     We provide metallicities for 81 stars, which cover the wide [Fe/H]=$-$3.2
     to $-$1.5 dex range. The abundances of ten other elements are derived: Mg,
     Ca, Ti, Sc, Cr, Mn, Co, Ni, Ba, and Eu.  Despite its small mass, Sextans is
     a chemically evolved system, showing evidence of a contribution from
     core-collapse and Type Ia supernovae as well as low-metallicity asymptotic
     giant branch stars (AGBs). This new FLAMES sample offers a sufficiently
     large number of stars with chemical abundances derived with high accuracy
     to firmly establish the existence of a plateau in [$\alpha$/Fe] at $\sim
     0.4$ dex followed by a decrease above [Fe/H]$\sim-2$ dex. These features
     reveal a close similarity with the Fornax and Sculptor dSphs despite their
     very different masses and star formation histories, suggesting that these
     three galaxies had very similar star formation efficiencies in their early
     formation phases, probably driven by the early accretion of smaller
     galactic fragments, until the UV-background heating impacted them in
     different ways. The parallel between the Sculptor and Sextans dSph is also
     striking when considering Ba and Eu. The same chemical trends can be seen
     in the metallicity region common to both galaxies, implying similar
     fractions of SNeIa and low-metallicity AGBs.  Finally, as to the iron-peak
     elements, the decline of [Co/Fe] and [Ni/Fe] above [Fe/H]$\sim -2$ implies
     that the production yields of Ni and Co in SNeIa are lower than that of
     Fe. The decrease in [Ni/Fe] favours models of SNeIa based on the explosion
     of double-degenerate sub-Chandrasekhar mass white dwarfs.}

\maketitle

\section{Introduction}

A large variety of topics rely on our understanding of the formation and
evolution of dwarf galaxies.  In the $\Lambda$CDM cosmology, galaxies form
hierarchically by cooling and condensation of the baryons within dark-matter
haloes that gradually merge \citep{White1978}. Strong efforts are therefore
concentrated on counting, characterising, and quantifying the impact of the earliest
and smallest of these galactic systems from the local universe to the
reionisation period \citep[e.g.][]{Tolstoy2009, Wise2014, Sawala2016,
  Simon2019, Torrealba2019}.

The Sextans dwarf spheroidal galaxy (dSph) was discovered by
\citet{Irwin1990}. At a distance of $\sim$ 90 kpc, it is one of the closest
satellites of the Milky Way \citep{Mateo1995, Lee2003, Battaglia2011}. Its late
discovery is the consequence of its large extent on the sky with a tidal radius
of $120 \pm 20$ arcmin \citep{Cicuendez2018} and low surface brightness of
$\sigma_0 = 18.2 \pm 0.5$ mag/arcmin$^{2}$ \citep{IrwinHatzidimitriou1995} making it a challenging galaxy to characterise given the large fraction of
Milky Way interlopers.

The analysis of the colour magnitude diagram (CMD) of Sextans reveals a stellar
population which is largely dominated by stars older than $\sim$11Gyr
\citep{Lee2009}, with evidence for radial metallicity and age gradients, the
oldest stars forming the most spatially extended component \citep{Lee2003,
  Battaglia2011, Okamoto2017, Cicuendez2018}.

Spectroscopic follow-up in Sextans started in 1991 at medium--low resolution in
the region of the calcium triplet (CaT, 8498, 8542 and
8662 \AA). \citet{DaCosta1991} identified six galaxy members and derived a mean
metallicity of [Fe/H] $= -1.7 \pm 0.25$ dex. \citet{Suntzeff1993} increased the
sample of galaxy members up to 43 and revised the galaxy peak metallicity to $-2.05
\pm 0.04$ dex. The enhanced multiplexing power of a new generation of spectrographs with over $\sim$ 30\arcmin\
fields of
view opened up the possibility to
analyse hundreds of stars at once. Spectroscopy in the region of the Mg I triplet
(5140-5180 \AA) and the CaT absorption features were used for rough chemical
tagging and to investigate the mass profile of  Sextans, as well as the possible existence
of kinematically distinct stellar populations at its centre
\citep{Walker2006a, Walker2009a, Battaglia2011}. Sextans is thought to be about
0.4 Gyr away from its pericentre (r$_{\rm{peri}}$ $\sim$ 75 kpc) moving towards
its apocentre (r$_{\rm{apo}}$ $\sim$ 132 kpc), and its orbit seems inconsistent
with a membership to the vast polar structure of Galactic satellites
\citep{Casetti-Dinescu2018, Fritz2018}. Until recently, no statistically
significant distortions or signs of tidal disturbances had been found down to very
low surface brightness limit \citep{Cicuendez2018}, but combined reanalysis
of the spectroscopic membership and deep photometry have revealed the presence of a
ring-like structure that is interpreted as the possible sign of a merger
\citep{Cicuendez2018b}.

One thread of studies traces the formation and evolution of galaxies by
exploring their chemical evolution as preserved in stellar abundance patterns.
Comparison between galaxies of very different star formation histories also
offers important insight into poorly understood nucleosynthetic origins such as
for the neutron capture elements
\citep[e.g.][]{Tolstoy2009,Jablonka2015,Mashonkina2017,Ji2019,Reichert2020}. Here,
spectroscopic multiplex again plays a fundamental role allowing to switch from
the pioneer ensembles of a few stars per galaxy that had elemental abundances
and abundance ratios \citep[e.g.][]{Hill1995,Shetrone2001, Shetrone2003} to
statistically significant samples.  A step forward arose from Keck/DEIMOS
medium-resolution (R$\sim$7000) spectroscopy, with about 35 Sextans stars with
delivered abundances of $\alpha$-elements (Mg, Si, Ca, Ti) at accuracies better
than 0.3 dex \citep{Kirby2011a}. This sample has recently been completed with
Cr, Co, and Ni abundances \citep{Kirby2018}. The number of remaining open
questions in Sextans is nevertheless very large. In particular, its low
metallicity range ([Fe/H]$\le -2.$) is still uncovered. Only a few extremely
metal-poor stars have been targeted in this galaxy \citep{Aoki2009, Tafelmeyer2010}. The mean
trend and scatter at fixed [Fe/H] of the abundance ratios still need to be
investigated over the full chemical evolution of Sextans.

The VLT/FLAMES fibre-spectrograph has already been transformative in addressing
similar questions in other dSphs. The Dwarf Abundances and Radial velocity team
(DART) targeted the Fornax and Sculptor dSphs \citep{Letarte2010,
  hill2019}. The present paper presents the DART high-resolution spectroscopic sample
of Sextans. In the following, we present the analysis of a sample of 81 Sextans
stars that have been observed in the central region of Sextans at high
resolution (R$\sim$20,000) with VLT/FLAMES (GIRAFFE and UVES). This is the
largest sample of red giant branch (RGB) stars dedicated to a chemical analysis, and allows us to address questions related to both stellar nucleosynthesis and galaxy
evolution.

\section{Observations and data reduction}

\begin{figure}
   \centering
   \includegraphics[width=\hsize]{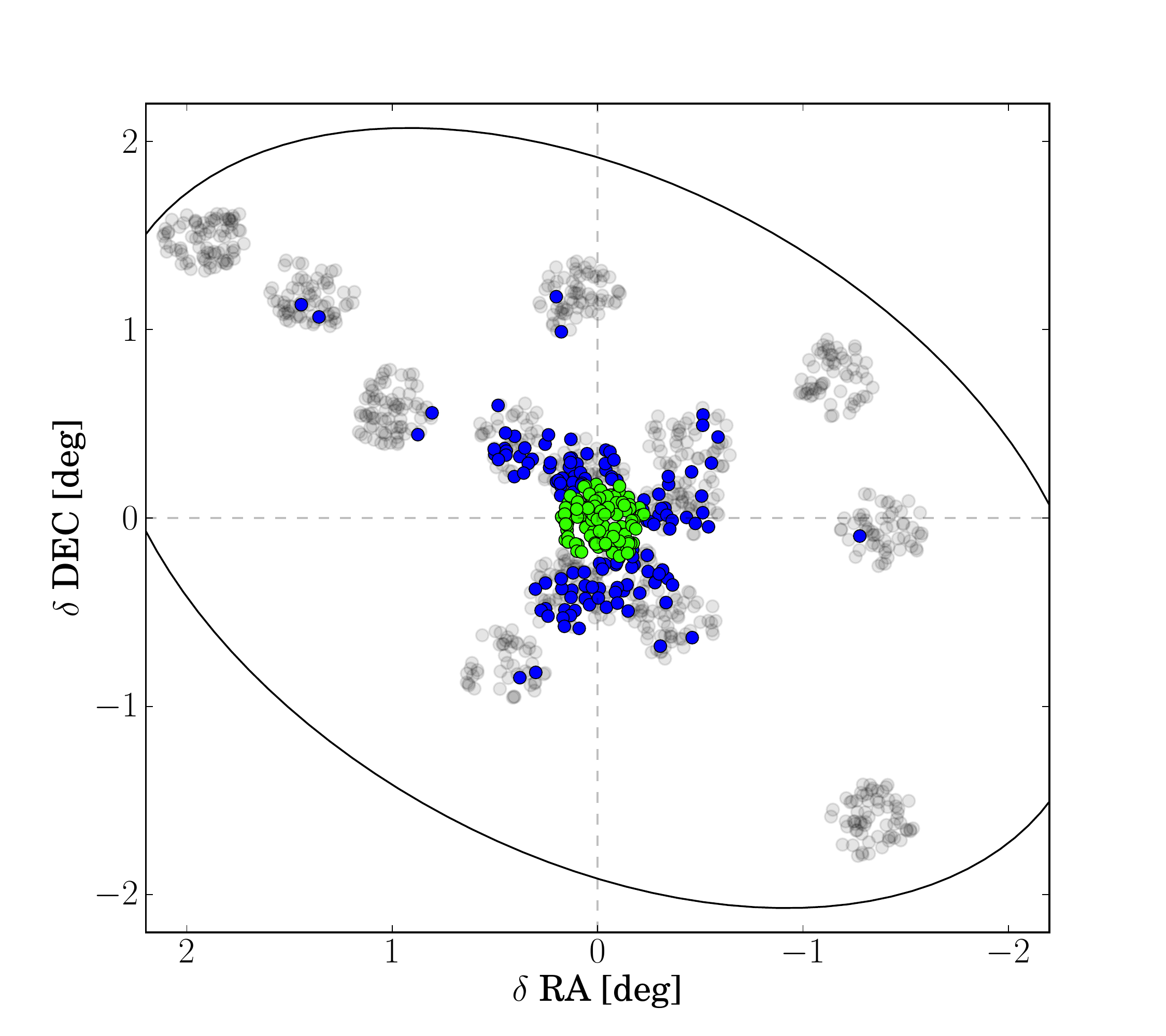}
   \caption{Spatial distribution of the DART spectroscopic
     observations in Sextans. The grey points show the stars
     observed around the CaT survey \citep{Battaglia2011}. The blue circles
     correspond to the probable members of this medium-resolution
     sample, while the green points are the stars of our high-resolution
     sample (see Table \ref{Photo_param} for the values of RA and
     DEC).  The black ellipse represents the tidal radius of Sextans
     calculated by \cite{Cicuendez2018}.}
         \label{Star_Location}
\end{figure}

\begin{figure}
  \centering
  \includegraphics[width=\hsize]{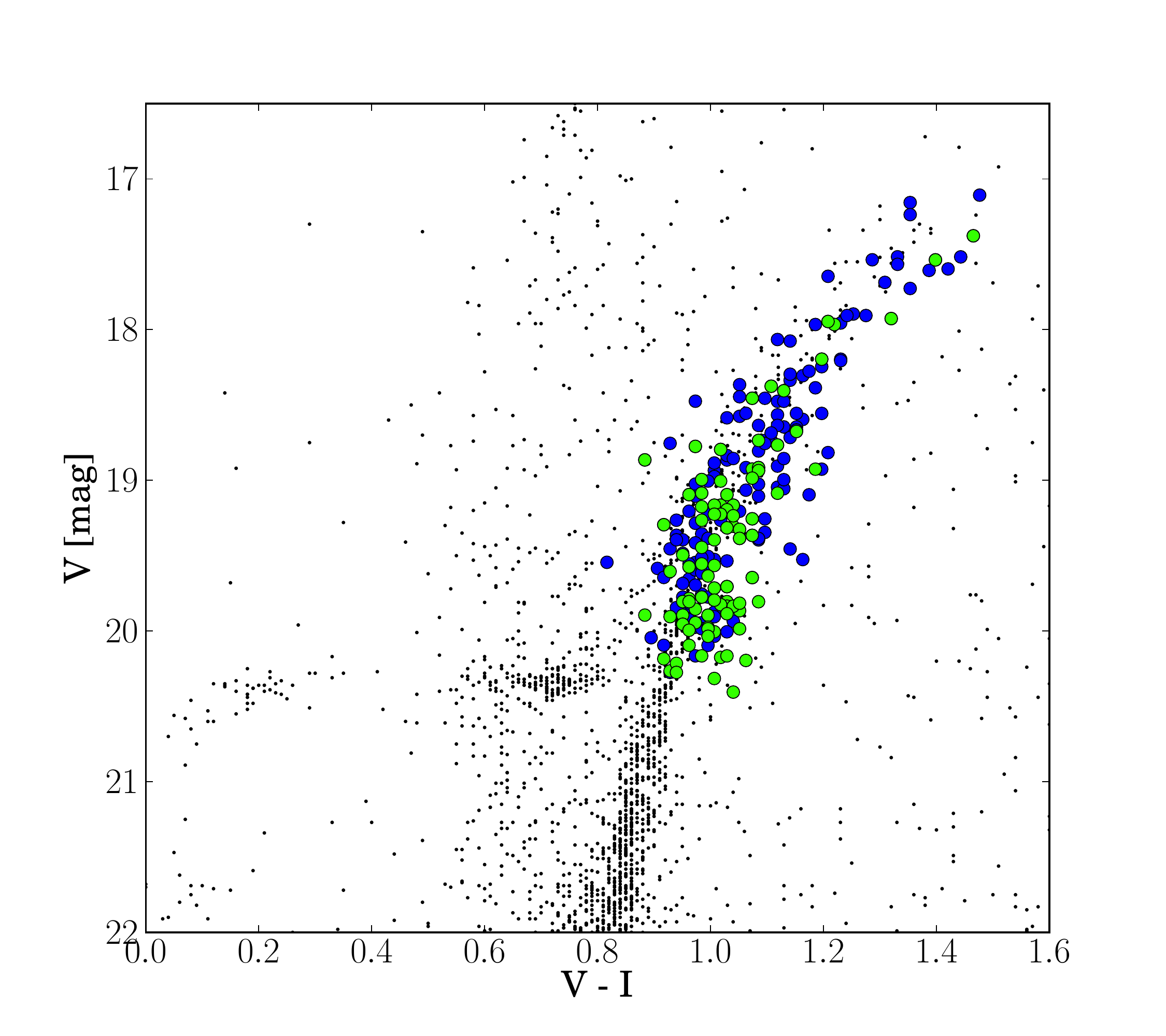}
      \caption{Colour-magnitude diagram of the Sextans stars. The blue and green points are in as in Fig. \ref{Star_Location}.
      The black dots correspond to Sextans photometry taken with the CFH12K CCD camera at Canada-France-Hawaii
      Telescope  \citep[CFHT;][]{Lee2003}.}
         \label{CMD}
\end{figure}

Our targets are RGBs located in the central
25\arcmin\  field of the Sextans dSph. About half of the sample is
composed of RGBs previously identified as Sextans members based on their radial
velocities measured from medium-resolution spectra around the calcium triplet
\citep{Battaglia2011}. The rest of the sample was selected from the
position of the stars in the CMD of Sextans. The
spatial distribution of our sample stars is presented in
Fig. \ref{Star_Location}, while Fig. \ref{CMD} displays their location along the
RGB $V$ versus $V-I$ plane.

We gathered high-resolution (R$\sim$20'000) spectra of 101 stars in the HR10,
HR13, and H14 gratings of the multi-fibre spectrograph FLAMES/GIRAFFE installed at
the VLT (ESO Program 171.B-0588(A)). The observations were conducted in three runs
from March to December 2004 and led to a total exposure time of 30 hours and 17
minutes. Two fibres were linked to the red arm of the UVES spectrograph, yielding
R $\sim$ 47 000 spectra of two stars, S05-5 and S08-229, over the 
$\lambda$ $\sim$ 4800-6800 \AA\ range. Table \ref{obs} summarises the characteristics
of the gratings and the corresponding exposure times.

In each MEDUSA plate, 16 (in HR10) and 19 (in HR13, HR14) fibres were
dedicated to the sky background. Five fibres were allocated to
simultaneous wavelength calibration in HR10 and HR13, but not in HR14
to prevent pollution from the argon lines which saturate at wavelength
longer than 6500 \AA. Nevertheless, in some cases these
wavelength calibration spectra have also polluted their closest
neighbour stellar spectra in HR10 and HR13.  The spectra of 25 stars are
affected by this problem; they are indicated in Table
\ref{Photo_param}. For those stars, we discarded the polluted spectra
(HR10 and/or HR13) from further abundance analysis. 
Three more stars have one or two missing parts of their spectra: S05-70
and S05-78 have no HR10 spectrum and S08-274 has only a HR10 spectrum. This
information is also reported in Table \ref{Photo_param}.

\begin{table}
\small
\centering                          
\begin{tabular}{lccc}
 \hline
 Grating & HR10 & HR13 & HR14\\
 \hline
$\Delta \lambda$  &  $\sim$5340-5620 \AA  & $\sim$6120-6400 \AA  & $\sim$6300-6700 \AA\\
 Resolution & 19800  &  22500  &  17740 \\
 Exposure Time &  36080 s &  43566 s  & 29400 s\\
\hline
\end{tabular}
\caption{Summary of the observations: The three gratings of the GIRAFFE observations are indicated as well as their wavelength coverage ($\Delta \lambda$)  and total exposure times.
Two fibres were simultaneously connected for the total exposure time to the red arm of UVES with a central wavelength of $\sim$5800~$\AA$ and a resolution of 47000.}             
\label{obs}      
\end{table}

The data reduction has been performed with the ESO GIRAFFE Pipeline version
2.8.9. For each science frame, we used the corresponding bias, dark, and
flat-field frames, as well as arc lamp spectrum.  For the sky subtraction, we
used the routine of M. Irwin \citetext{priv.\ comm.}, which creates an average sky
spectrum from the sky fibres. This average sky spectrum was then subtracted from
each target spectrum after appropriate scaling in order to match the sky
features in each fibre. We extracted the individual spectra with the task
\textit{scopy} in \verb+IRAF+.  The heliocentric velocities were calculated and
possible shifts were corrected before the individual subexposures were averaged
with \textit{scombine} in \verb+IRAF+ with an averaged sigma clipping algorithm that removes remaining cosmic rays and bad pixels.

The signal-to-noise ratio (S/N) per pixel of each of our 101 spectra was estimated
with the \verb+IRAF+ task \textit{splot} in three continuum regions: [5456.6\AA
  , 5458.94\AA], [6193.47\AA , 6197.74\AA], and [6520.0\AA , 6524.4\AA] for
HR10, HR13, and HR14, respectively.  The results are presented in
Table~\ref{Atmospheric_param_from_photometry}.

The spectra were normalised with the routine of M. Irwin \citetext{priv.\ comm.}, which consists in an iterative process of spectrum filtering. The routine detects
the lines through asymmetric k-sigma clipping and they are masked to fit the
continuum.

\section{Selection of the Sextans members from the GIRAFFE sample}

\begin{figure}
  \centering
  \includegraphics[width=\hsize]{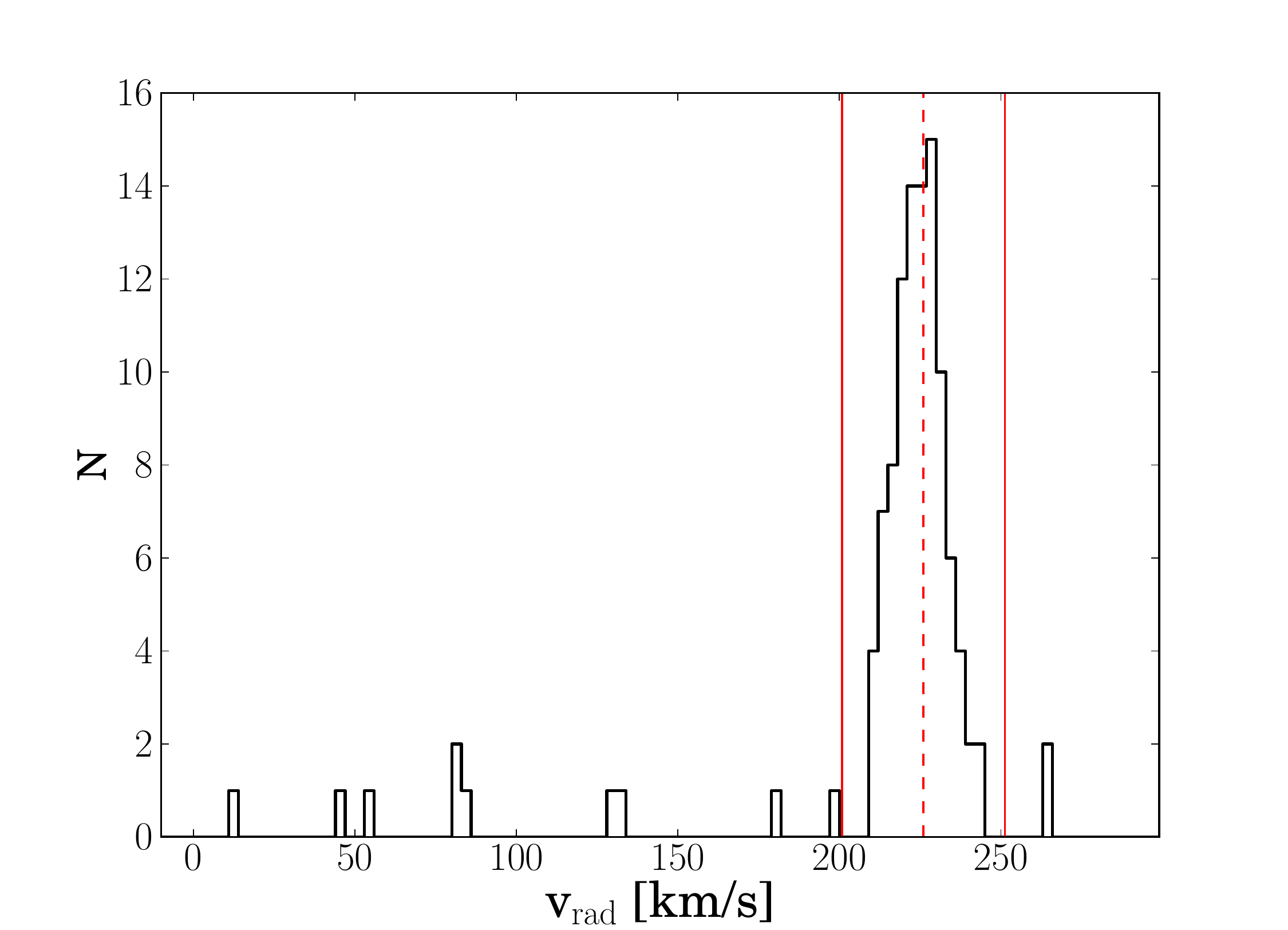}
  \caption{Radial velocity distribution of our stars (black solid line). The
    dashed red line corresponds to the mean radial velocity from
    \citet{Battaglia2011} based on 174 probable members: $v_{sys, \ helio} =
    226.0 \pm 0.6$ km/s. The two solid red lines represent the $\pm 3 \sigma$
    interval within which the membership zone is defined.}
  \label{Vrad_histo}
\end{figure}

We derived the stellar radial velocities from the 1D reduced spectra with
\verb+DAOSPEC+ \footnote{DAOSPEC has been written by P.B. Stetson for the
  Dominion Astrophysical Observatory of the Herzberg Institute of Astrophysics,
  National Research Council, Canada.}  \citep{StetsonPancino2008}, which
cross-correlates all the detected lines with an input line list that we took as
in \citet{Tafelmeyer2010}.  The radial velocities in each grating (HR10, HR13,
HR14) and their mean are provided in Table \ref{vrad_HR} and shown in Figure
\ref{Vrad_histo}.

The Sextans members were selected by requiring that their radial velocities fall within 3$\sigma$ of the galaxy systemic velocity derived by
\citet{Battaglia2011}, $226.0 \pm 0.6$ km/s, $\sigma = 8.4 \pm 0.4$ km/s.
Twelve stars were identified as background or foreground interlopers; they are
indicated as \textit{Non-member (}v\textit{)} in Table \ref{Photo_param}. Our
final cleaned sample has a mean velocity and dispersion of v$_{\rm sys, \rm
  helio} = 224.9 \pm 1.9 $ km/s, $\sigma = 7.7 \pm 0.5$ km/s in perfect
agreement with \citet{Battaglia2011}.

The dataset has been acquired over a time period of 1 month, and this time delay between
the HR10 and HR13/HR14 spectra allowed us to look for binaries.  These  were
identified based on radial velocities derived in each grism differing by more
than 1$\sigma$, the error on the velocity. This corresponded to differences in
velocities of $\Delta$v$_{\mathrm{rad, HR}}$ between 7 and 35 km/s.  A total of
$13$ binary systems were identified in this way, for which the mean radial
velocities in Table \ref{Photo_param} are preceded by a tilde symbol. We also
identified five carbon stars with a strong C$_2$ band head. A preliminary analysis shows
that these are CEMP-s stars. Further investigation of these stars is postponed to another
paper. 

Five stars (S05-72, S08-321, S08-111, S08-301, and S08-293) with overly poor-quality
spectra and very unstable solutions for their abundances have also been removed
from our investigation. These latter stars usually fall at the margins of the photometric
RGB and are classified as \textit{Non-member (c)} in Table
\ref{Photo_param}.
As a result of the above selection, our final sample encompasses 81 RGB Sextans stars.

\section{Stellar atmosphere models}

Our full sample is distributed along the Sextans RGB down to the horizontal branch
level, that is, reaching uncommonly faint magnitudes for high-resolution
spectroscopic analyses.  For that reason, the elemental abundances of the GIRAFFE
sample were derived by synthesis based on a $\chi^{2}$ minimisation procedure,
which can balance the derivation of the chemical abundances over the widest
possible wavelength range at once. This procedure is described in
Sect. \ref{synth}.  Two stars were sufficiently bright to be fed to
UVES. These have been analysed in a classical way, which is detailed in Sect. \ref{uves},
based on the equivalent widths of their individual absorption lines.

In both cases, we adopted the MARCS 1D spherical atmosphere models, which were
downloaded from the MARCS web site\footnote{marcs.astro.uu.se}
\citep{Gustafsson2008}, and interpolated using the code of Thomas Masseron available
on the same website. We assumed standard values of [$\alpha$/Fe] following the
Galactic disc and halo, namely +0.4 for [Fe/H]$\leq$ $-1.5$, $+0.3$ for
[Fe/H]=$-0.75$, $+0.2$ for [Fe/H]=$-0.54$, $+0.1$ for [Fe/H]=$-0.25,$ and $+0.0$
for [Fe/H] $\geq$ 0.0.

\section{Photometric parameters}

The photometric estimates of the stellar parameters \teff, \logg, and \vt\ were
based on the ESO 2.2m WFI $V$ and $I$-band magnitudes \citep{Tolstoy2004,
  Battaglia2006, Battaglia2011}, as well as on the $J$, $H$, $Ks$ WFCAM LAS UKIRT
photometry calibrated onto the 2MASS photometric system
\citep{Skrutskie2006}. These magnitudes are provided in Table \ref{Photo_param}.

For each star, the photometric effective temperature is taken as the
simple average of the four colour temperatures $T_{\rm V-I}$, $T_{\rm
  V-J}$, $T_{\rm V-H}$, and $T_{\rm V-K}$ obtained with the
calibration of \citet{Ramirez2005}.  When possible, the metallicity
estimate from the CaT was used as initial iron abundance; otherwise we
adopted the mean metallicity of the Sextans stellar population,
[Fe/H]=$-1.9$, as given by \citet{Battaglia2011}.  We used $E_{B-V}
= 0.0477$ \citep{Schlegel1998} and the reddening law, $A_{V} =
3.24\ E_{B-V}$, of \citet{cardelli1989}.  The stellar surface
gravities were estimated from the stellar effective temperature and
the bolometric correction calculated as in \citet{Alonso1999}. We
adopted a distance of 90 kpc \citep{Tafelmeyer2010, Battaglia2011}, a
0.8 M$_{\sun}$ stellar mass for our sample stars and
\logg$_\sun=4.44$, \teff$_\sun$ = 5790 K, and M$_{Bol, \sun} = 4.75$.
The micro-turbulence velocities were derived according to the
empirical relation \vt = 2.0 $-$ 0.2 $\times$ \logg\ of
\citet{AnthonyTwarog2013}. The effective temperature for each colour,
the bolometric corrections and initial metallicities are provided in
Table \ref{Atmospheric_param_from_photometry}.

These photometric parameters were the final ones for the GIRAFFE sample, while
\teff\ and \vt\ could be further spectroscopically adjusted for the two stars
observed with UVES as described in Sect. \ref{uves}.

\section{Analysis of the two stars observed with UVES}
\label{uves}

S05-5 and S08-229 were sufficiently bright to be allocated to two of the eight
UVES fibres of the FLAMES configuration. These stars were analysed in the same
way as in our previous publications on high-resolution spectroscopic studies in
dwarf spheroidal galaxies \citep[e.g.][]{Letarte2006, Letarte2010,
  Tafelmeyer2010, Jablonka2015, hill2019}.  The abundance analysis has been
performed using the local thermodynamical equilibrium (LTE) code \textit{CALRAI}
first developed by \citet{Spite1967} (see also \citet{Cayrel1991} for the atomic
part), and continuously updated over the years. We summarise the main steps
below.

\subsection{Measurements}
\label{sec:eqw}

The absorption features were measured following the line list of
\citet{Tafelmeyer2010} which combines those of \citet{Letarte2010} and
\citet{Cayrel2004}. The equivalent widths were measured with
DAOSPEC. As DAOSPEC fits absorption lines with Gaussians
that have a fixed FWHM, the equivalent widths of strong lines with prominent wings
are systematically underestimated.  Therefore, lines with equivalent widths larger
than 180~m$\AA$ were discarded. Table \ref{tab:uves_eqw} lists the lines and their equivalent widths.
The present analysis focuses on the elements derived in the HR10, HR13, and
  HR14 grisms. A discussion of the other elements accessible thanks to the
  UVES wavelength coverage and high resolution is deffered to a forthcoming
  paper.

\subsection{Final atmospheric parameters and error estimates}
\label{sec:specparams}

The convergence to our final effective temperatures and the microturbulence
velocities (\vt) presented in Table \ref{Stellar_param} was achieved
iteratively, as a trade off between minimising the trends of metallicity derived
from the \ion{Fe}{I} lines with excitation potentials and equivalent widths on the one
hand and minimising the difference between photometric and spectroscopic
temperatures on the other hand.  Starting from the initial photometric
parameters we adjusted \teff\ and \vt\ allowing for deviation by no more than 2
$\sigma$ (the uncertainty) of the slopes. This yielded new metallicities which
were then fed back into the photometric calibration to get new photometric
temperatures and gravities.  No more than two or three iterations were needed to
converge to our final atmospheric parameters.

The predicted equivalent widths rather than the observed ones were considered in
this procedure, because the errors on the measurements can bias the slope of the
diagnostic plots \citep[see][]{magain84}. However, the surface gravities were kept
fixed to their photometric values due to the small number of \ion{Fe}{II} lines
and possible NLTE effects, which might affect the ionisation equilibrium.

The final abundances are calculated as the weighted mean of the abundances
  obtained from the individual lines, where the weights are the inverse
  variances of the single line abundances. These variances were propagated by
  \textit{CALRAI} from the estimated errors on the corresponding equivalent
  widths. They are listed in Table \ref{tab:UVES_abund}. Table \ref{tab:atmoerr} provides the errors on the abundances linked to the
  uncertainties on atmospheric parameters for the two stars observed with UVES.

\section{Analysis of the GIRAFFE sample}
\label{synth}

\subsection{Synthetic grid}

\noindent {\bf [Fe/H]:} To determine [Fe/H], a library of synthetic spectra covering the HR10, HR13, and
HR14 wavelength regions was created with \verb+MOOG+ \footnote{ \texttt{MOOG}
  \citep{Sneden1973} is a \texttt{FORTRAN} code designed to perform a variety of
  LTE line analyses and spectrum synthesis with the aim being to help with determination of the stellar
  chemical composition. The basic equations of the stellar line
  analysis in LTE are used following the formulation of Edmonds (1969).  Much of
  the \texttt{MOOG} code follows in a general way the \texttt{WIDTH} and
  \texttt{SYNTHE} codes of R. L. Kurucz (http://kurucz.harvard.edu/).} (August
2010 Version). The stellar spectra were generated at the resolution R=40000,
over  4200K $\leq$ \teff $\leq$ 5300K  with a step of 50
K,   0.5 $\leq$ \logg $\leq$ 2.5 with a step of 0.1 dex,
 $1.4 \leq$ \vt $\leq 2.0$ with a step of 0.2 km/s,
and  $-4.0 \leq $[Fe/H] $ \leq -0.5$ with a step of
0.1 dex.\\



\noindent {\bf Elemental abundances:} To derive the other elemental abundances,
          [X/H], we created two sets of synthetic spectra at given
          $T_\mathrm{eff}$, \ensuremath{\log g}, v$_{\rm turb}$, and [Fe/H]. The
          first set of synthetic spectra covered an interval of 6 dex in [X/H],
          from [X/H] $= -5.0$ to [X/H] $= 1.0$ , in $0.5$ dex steps. The second
          set covered only 2 dex in [X/H] but with steps of $0.05$ dex and was
          centred on the first estimated [X/H] of the star under analysis. \\

The resolution of the synthetic spectra was adjusted to those of the observed
spectra, i.e. to resolving powers $R=19800$ for HR10, $22500$ for HR13, and
$17740$ for HR14, by convolution with Gaussians with FWHM=$15.2$, $13.3$, and
$16.9$~\kms, and was then normalised with \verb+MOOG+. A number of different
macroscopic mechanisms can broaden the stellar absorption features. The dominant
one in our case is the instrumental broadening. However, as is often the case,
applying this instrumental broadening to theoretical spectra is insufficient.
Macro-turbulence and possible rotation need to be taken into account.  These
corrections were estimated on the four brightest stars of the GIRAFFE sample for
which both a classical analysis based on line equivalent widths and the spectral
synthesis could be independently performed (see Sect. \ref{eqwsynth}). For this
particular purpose, we only considered the \ion{Fe}{I} lines. We found that an
additional broadening by $\sigma$=9.0 km/s, 7.9 km/s and 7.6 km/s Gaussians for
the HR10, HR13, and HR14 gratings, respectively, resulted in the best agreement
in metallicities between the two types of analyses.  These broadenings were
therefore applied to the synthetic spectra for the three respective setups.

\subsection{Synthesis principles}

The elemental abundances of the GIRAFFE sample stars were  derived by
synthesis based on a $\chi^{2}$ minimisation procedure, in which
\begin{equation}
\chi^{2} = \sum\limits_{i=1}^N  \dfrac{(y_{\rm obs, i} - y_{\rm mod, i})^{2}}{y_{\rm mod, i}} 
 \end{equation}

where $i$ is the pixel index and $y$ the normalized flux of this pixel in the
observed $y_{\rm obs, i}$ spectrum as well as in the model synthetic $y_{\rm mod,
  i}$ one.

For each chemical species, a global $\chi^{2}$ was calculated for all detected
lines. The match between the synthetic and observed spectra was estimated in
windows centred on the lines with widths of 22 pixels in HR10 and HR13 (1.05\AA), and
28 pixels (1.35\AA) in HR14. The width of these windows were calculated as 1.75
$\times$ FWHM, with FWHM being the spectral resolution of the grating under
consideration.

The final $\chi^{2}$ value was summed over the lines in the three wavelength bands :  
\begin{equation}\label{eqn:chi2_tot}
\chi^{2}_{\rm tot} = \left( \sum\limits_{\rm HR10} \chi^{2}\right) \ + \left(\sum\limits_{\rm HR13} \chi^{2}\right) \ + \left(\sum\limits_{\rm HR14} \chi^{2}\right)
\end{equation}.

\subsection{Iron abundance}

Starting from the \ion{Fe}{I} line list of \citet{Tafelmeyer2010}, we discarded
the weakest lines which would be overly altered by the noise or
blended. This selection ensures that our sample stars, spanning a large portion
of the RGB, are all analysed in a homogeneous way. Our final line list is
presented in Table \ref{Linelist}. It encompasses 50 \ion{Fe}{I} lines: 20 in
HR10, 21 in HR13, and 9 in HR14. The HR13 and HR14 grisms have a small
overlapping region. The lines in common were considered in the HR13 part only,
taking advantage of its higher resolution.

For each star of photometric \teff, \logg, and \vt, we selected the closest
synthetic spectra with these parameters, namely with \teff\ within 50 K and
\logg\ within 0.1 dex. We interpolated them at the value of \vt.  The
$\chi^{2}$ values were then calculated for [Fe/H] varying from $-4.0$ to $-0.5$
in steps of 0.1 dex. Finally, the $\chi^{2}$-[Fe/H] relation was interpolated to
a $0.01$~dex step level in order to properly locate the minimum of the relation.  The
corresponding metallicities are given in Table \ref{Stellar_param}.

We note that the maximum differences in \teff\ and \logg\ between the
photometric stellar parameters and the values adopted in the grid are
$25$~K and $0.05$~dex, respectively. These differences are smaller than the
systematic errors on the photometric estimates.

Re-inserting the final value of [Fe/H] into the colour--temperature
formula does not significantly affect the effective temperatures. The
maximum difference $\Delta T_\mathrm{eff}$ with respect to the initial
estimate is $45$~K, which is twice smaller than the systematic errors.
Therefore, we kept the initial effective photometric temperatures.

\subsection{Elemental abundances}

Once [Fe/H] was determined, we derived the abundances of the rest of the
elements using the same chi-squared minimisation procedure as for the
metallicity.  Again, the $\chi^{2}$-[X/H] relation was interpolated to reach a
[X/H] step of 0.01 dex.
Thanks to the wavelength coverage of our spectra we were able to
derive abundances of ten elements: Mg, Ca, Sc, Ti, Cr, Mn, Co, Ni, Ba, and Eu.

\begin{figure*}
\includegraphics[width=\hsize]{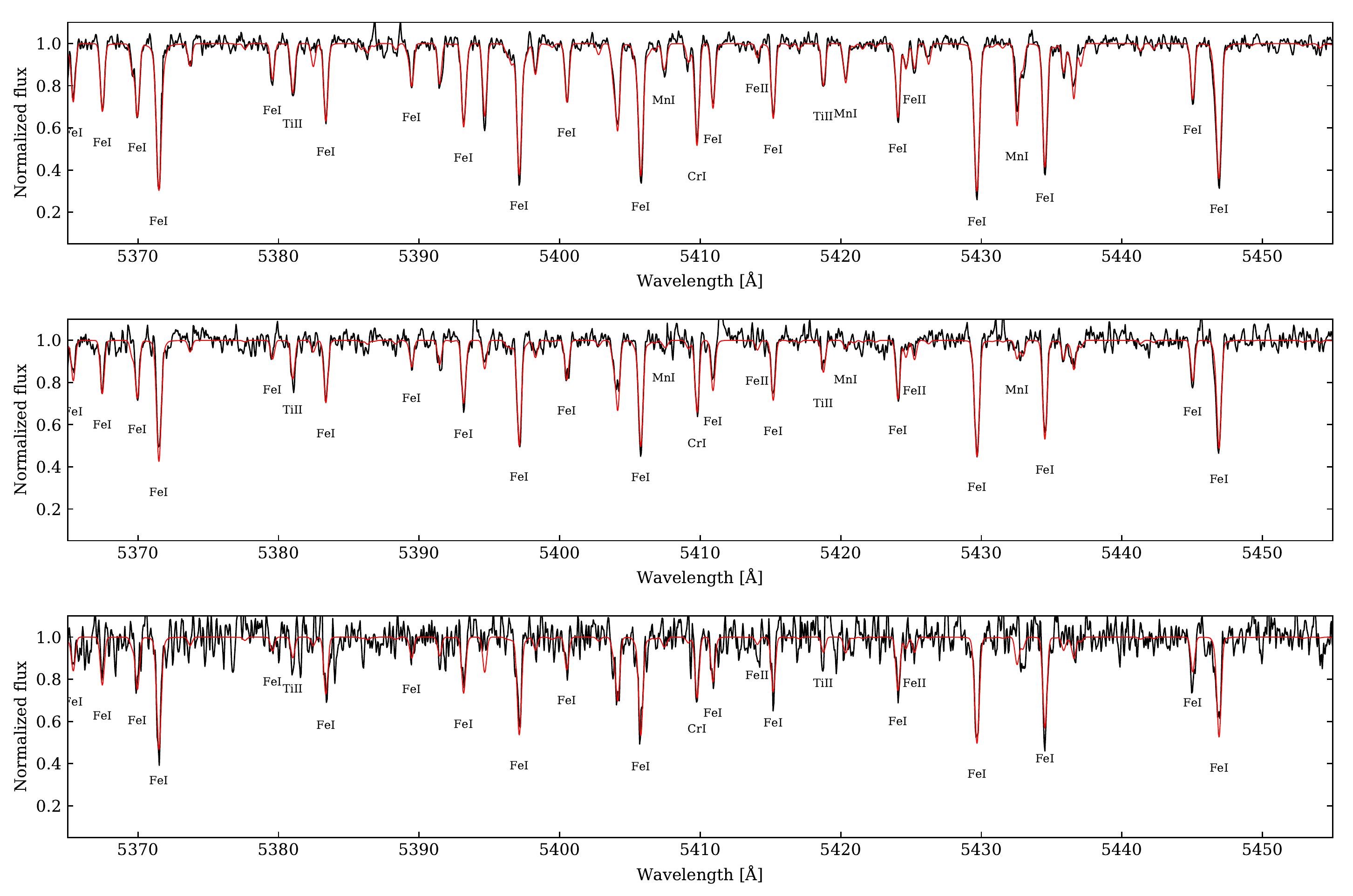}
\caption{Part of the HR10 spectrum of S08-6 (top), S08-242 (middle) and S05-67
  (bottom) in black with their synthetic best fits over-plotted in red.  A set
  of lines used in abundance determination are indicated for each star.  The
  stars are ranked from top to bottom by decreasing signal-to-noise ratios:
  $83$, $38$, and $23$.}
\label{spectra}
\end{figure*}

We took into account the hyperfine structure for the odd atomic number isotopes :
\ion{Sc}{II} \citep{Prochaska2000}, \ion{Mn}{I} (from Kurucz
database\footnote{available at
  http://kurucz.harvard.edu/linelists.html} as in \citet{North2012}),
\ion{Co}{I} \citep{Prochaska2000}, \ion{Ba}{II} \citep{Prochaska2000}, and
\ion{Eu}{II} (\citet{Lawler2001} as in \citet{VanderSwaelmen2013}).
The complete line list with wavelengths, excitation potentials,
oscillator strengths, and $C_6$ constants is presented in
Table~\ref{Linelist}.  The hyperfine components are indicated in
italics and their corresponding lines are followed by the text
`\textit{(equi)}'.  The abundances are given in Tables
\ref{Abundances_A}, \ref{Abundances_B}`, and \ref{Abundances_C}. The
solar abundances from \citet{GrevesseSauval1998} are repeated in the
first line of these three tables.
Figure \ref{spectra} presents three examples of observed and  synthetic best fits that span 
the range of S/N encountered in this study.

Figure \ref{spectra} presents three examples of observed and  synthetic best fits that span 
the range of S/N encountered in this study.

\subsection{Error budget}
The errors on the abundances derived for the GIRAFFE sample, reported as
error bars in all figures, correspond to the quadratic sum of the systematic
errors and the random errors defined below and indicated in Tables
\ref{Abundances_A}, \ref{Abundances_B}, and \ref{Abundances_C}.

\begin{itemize}
\renewcommand{\labelitemi}{$\bullet$}
\item Systematic errors

Taking as systematic errors a typical variation of $\pm$ 100 K in effective
temperatures, which accurately represents (75\% of our sample) the maximum variation
in \teff\ from one colour to the other, we re-derived the stellar parameters and
abundances for each star. We obtained small systematic shifts: between $0.04$
and $0.075$ for the  surface gravities, and an average of $0.01$~\kms on
the micro-turbulence velocities.  The systematic errors on [Fe/H] range from $0.1$
to $0.17$~dex, with an increasing trend with decreasing effective
temperature. For the other abundances, the smaller systematic errors are found
for Mg, Sc, Ti, and Eu with $0.05$~dex on average and the larger for Cr and Mn
with $0.2$~dex on average. Examples are provided in Fig.~\ref{Err_FeH_XFe}.

We also considered the effect of the uncertainty on the distance of
Sextans, which is $\sim \pm$ $5$~kpc around our adopted value \citep{Irwin1990,
  Lee2003}.  This translates into an uncertainty of $0.05$~dex on the surface
gravities, and corresponds to the same order of magnitude as
varying the effective temperatures by $\pm$ 100 K.\\

   \begin{figure}[!h]
   \centering
   \includegraphics[width=\hsize]{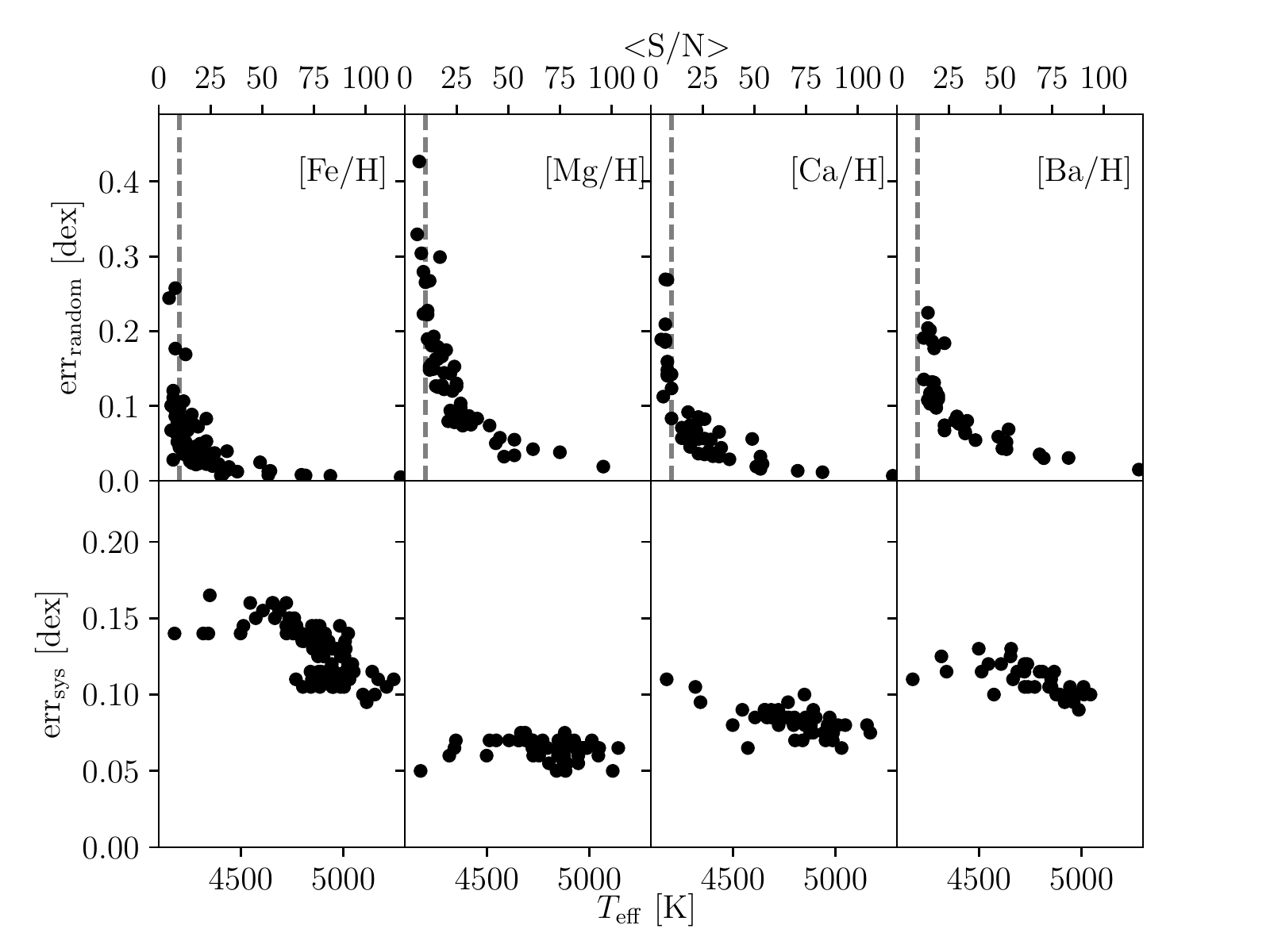}
   \caption{Distribution of the systematic and random errors as a function of S/N and effective temperature. These
     examples are shown for the elements for which we have the largest number of detections.
      We refer to the main text for a detailed definition of these errors.}
      \label{Err_FeH_XFe}
   \end{figure} 

\item Random errors

Variations in S/Ns may well impact the results
differently depending on the atmospheric parameters and the chemical
enrichment of the stars. Our sample spans more than 1 dex in [Fe/H]
and 1000 K in \teff.

The random errors on the abundances were therefore estimated from Monte Carlo
simulations. For each star, at given $T_\mathrm{eff}$, \ensuremath{\log g}, v$_{\rm turb}$, and [Fe/H], we generated
1000 spectra with the same S/N as the observed spectrum. The
abundances were then determined on these simulated noisy spectra. 
The adopted random error corresponds to the standard deviation of the Gaussian function that best fits the distribution of the  1000 obtained abundances.
Figure \ref{Err_FeH_XFe} presents some example of these randoms errors on [Fe/H] and [X/H] 
as a function of S/N.
From one set to the another, the higher the effective temperature and surface gravity, the
larger the impact of the noise, because the lines are smaller.

Based on this exercise, we only considered the abundances of
stars for which we have spectra of S/N $>$ 10  as secure. Indeed, regardless of the reliability of the metallicities, the errors on the other elements can reach $1$~dex,
hampering any reliable interpretation.

\end{itemize}

\section{How well do we perform ?}

The purpose of this section is to assess the quality of our analysis and also to
check that previously published stellar abundances in Sextans can be combined with our work
to provide a robust view of Sextans' chemical evolution.

\subsection{Comparison with the medium-resolution CaT sample}

\begin{figure}
\centering
\includegraphics[width=\hsize]{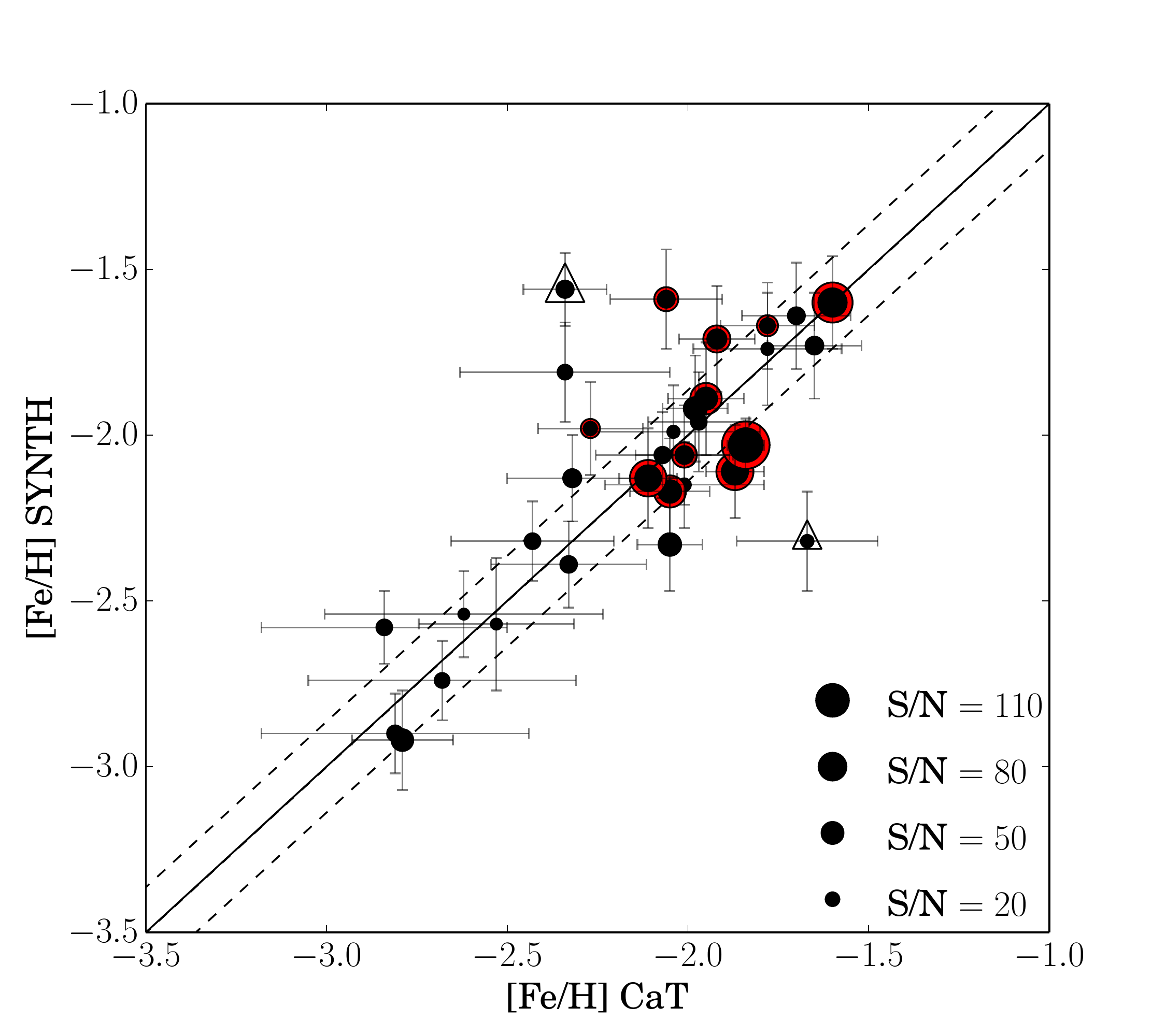}
\caption{Comparison between the CaT metallicity estimates and the
  [Fe/H] values derived from spectral synthesis (SYNTH). The size of the points
  is a function of the S/N of the spectra.  The two stars
  S08-75 and S08-59, which have been discarded from further analysis
  (see text) are identified with triangles. We highlight in red the
  stars which have been independently analysed with a classical method based on
  equivalent widths and which are in Fig.\ref{FeH_Fit_Eqw}.}
\label{FeH_CaT_vs_FeH_Synth}
\end{figure}

Calcium triplet metallicity estimates are available for a subset of 32 stars
with S/N $> 10$ \citep{Battaglia2011}. Figure \ref{FeH_CaT_vs_FeH_Synth}
presents a comparison between the [Fe/H] estimates of this latter study and
ours.  The agreement is excellent with a mean difference of $0.0014$~dex and a
standard deviation of $0.137$~dex. This comparison led us to discard from
further analysis two stars with $\Delta$[Fe/H]$> 0.4$ dex common to both
studies: S08-75 and S08-59, identified with triangles in
Fig. \ref{FeH_CaT_vs_FeH_Synth}.  A closer look indeed reveals that the
metallicity of S08-75 is based on only nine iron lines because its spectra have
been polluted by simultaneous calibration fibres in HR10 and HR13.  The position
of S08-59 on the CMD of Sextans sheds doubt on its RGB status; it is most likely
a horizontal branch star.

\subsection{Equivalent width versus synthesis}
\label{eqwsynth}

Given that the chemical abundances in the literature are generally derived with
methods based on equivalent widths, but also, and more specific to this work,
because two of our sample stars were gathered with UVES and analysed in this
classical way, it is important to check for the existence of any significant bias
between this type of analysis and the spectral synthesis (`SYNTH').

To this aim, a subsample of 11 bright (V$<$ 19 mag) stars (S08-3, S08-6,
S05-10, S08-38, S08-239, S08-241, S08-242, S07-69, S05-60, S08-246, and S08-183)
from the GIRAFFE sample was independently analysed with \textit{CALRAI}
in the same way as our two UVES stars. We refer to this method as ``EQW'' in the following.  The only difference with the analysis of the UVES
stars is that we fixed \teff\ to its photometric value. Only the turbulence
velocity \vt\ was determined by requesting a null slope (within the fit uncertainties
resulting from the measurement errors) of the relation between the abundances
and equivalent widths of the individual \ion{Fe}{I} lines. The line list and equivalent widths are
provided in Table \ref{tab:eqw}.

\begin{figure}
   \centering
   \includegraphics[width=\hsize]{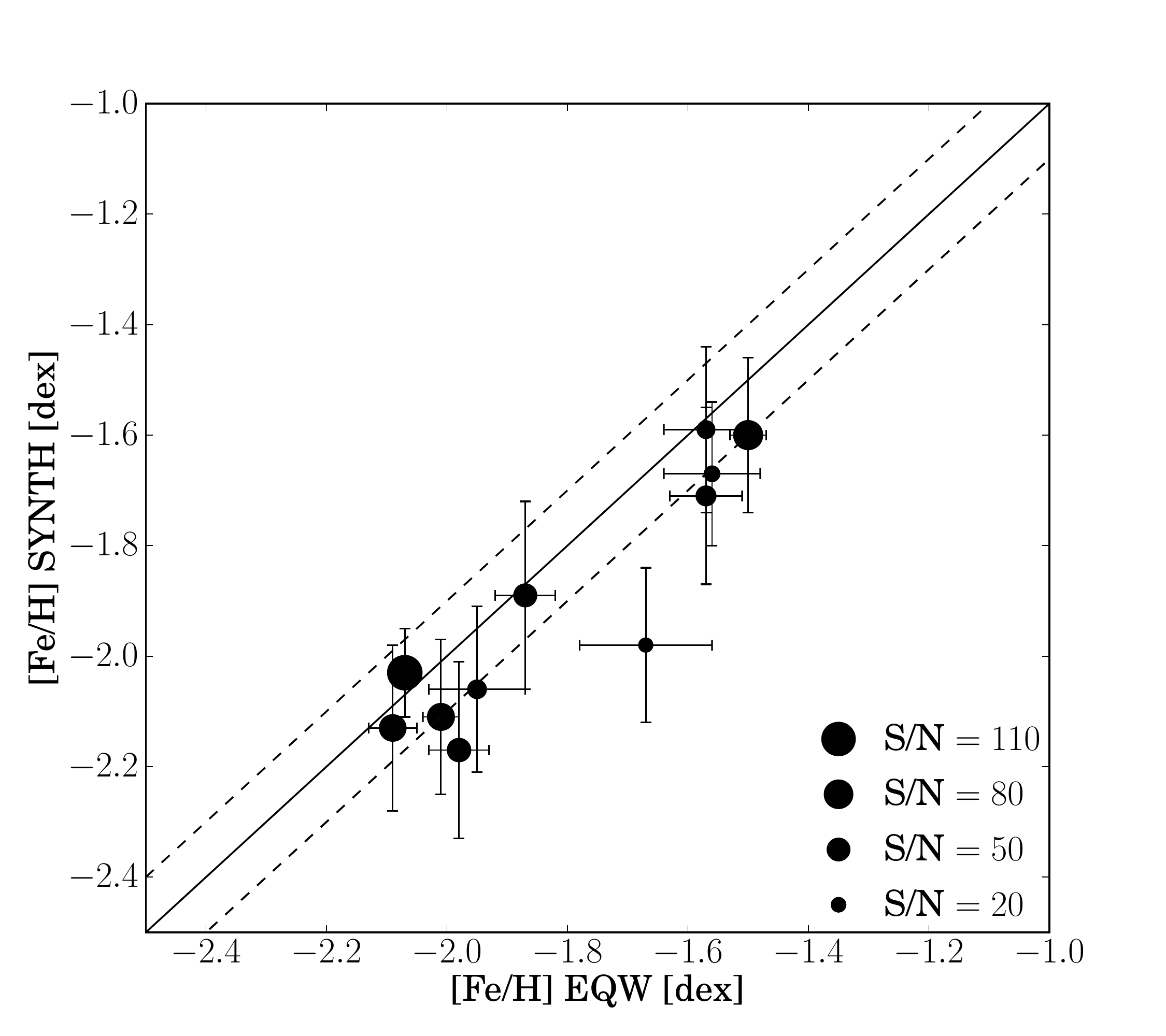}
      \caption{Comparison between the
        metallicity derived from our synthesis code, [Fe/H]$_{\rm SYNTH}$,
      and with a classical method based on the equivalent widths of the individual
      iron lines Fe/H]$_{\rm EQW}$ (see text). The solid line traces the line of
        equality while the dashed lines indicate a variation by $\pm 0.1$ dex. The
        size of the black circles increases with the S/N of the spectra.}
        \label{FeH_Fit_Eqw}
   \end{figure}

Figure \ref{FeH_Fit_Eqw} summarises the comparison between the two methods, EQW
and SYNTH, for the determination of \ion{Fe}{I}. The sizes of the symbols are proportional to the spectrum average S/Ns.  The dashed lines are
placed at $\pm 0.1$ dex of the $x=y$ line. There is a tendency for the EQW
\ion{Fe}{I} abundances to be slightly higher -- by $\sim$0.08 dex on average,
with a standard deviation of 0.09 dex -- than the SYNTH metallicities.  

\begin{figure}
   \centering
   \includegraphics[width=\hsize]{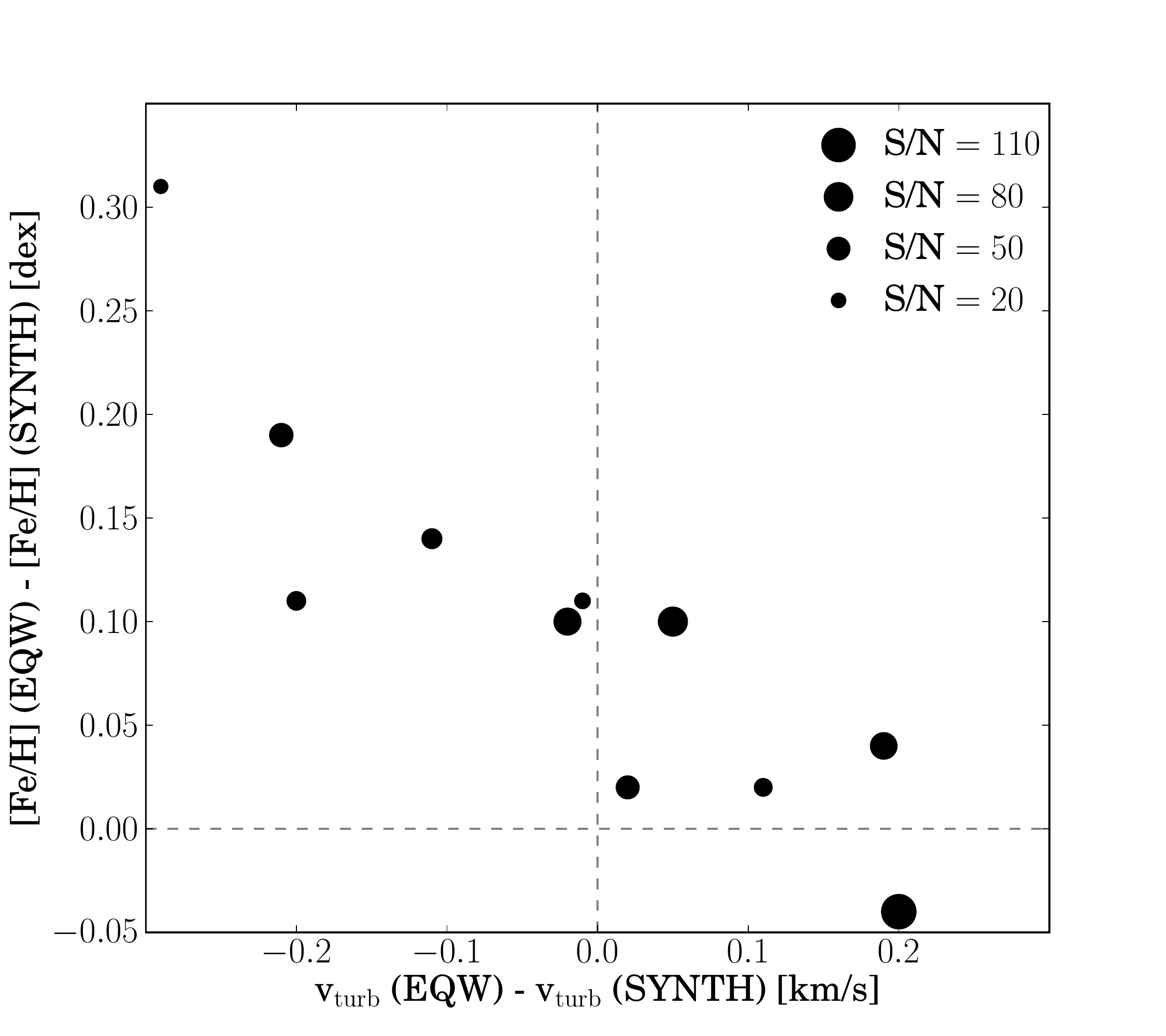}
      \caption{Differences between the metallicities obtained by the
        EQW and SYNTH methods as a function of the differences of the
        stellar micro-turbulence velocities. }
        \label{Delta_FeH_vs_Delta_vturb}
   \end{figure}

Even if this difference falls well within the error bars, it is interesting to
try and understand its origin.  Different factors can be invoked and they are
sometimes combined: a variety of S/Ns, the use of different
line lists, and different ways of deriving the micro-turbulence velocities
between the two techniques (requesting a null slope comparing \ion{Fe}{I}
abundances and equivalents widths for EQW, while it is estimated from an
empirical relation linked to \logg\ for SYNTH).  To illustrate this latter
point, the relation between the differences in [Fe/H] and the variations in \vt\
is shown in Fig.~\ref{Delta_FeH_vs_Delta_vturb}: a difference in \vt\ of
$0.1$~\kms can result in a $\Delta $[Fe/H] $\geq 0.1$ dex.  The star S05-60
has the largest metallicity difference between EQW and SYNTH. As a
matter of fact, the micro-turbulence velocities differ by $0.3$~\kms between the
two methods.

The case of the star S08-6 illustrates the influence of the linelist. The
difference of $0.1$ dex between the EQW and SYNTH results,
[Fe/H]$_{\rm EQW} = -1.5$ and [Fe/H]$_{\rm SYNTH} = -1.6$, is reduced
  to $0.03$~dex simply by using the exact same \ion{Fe}{I} line
  list. In this case, [Fe/H] $_{\rm SYNTH} = -1.53$ dex.

\begin{figure}
\centering
\includegraphics[width=\hsize]{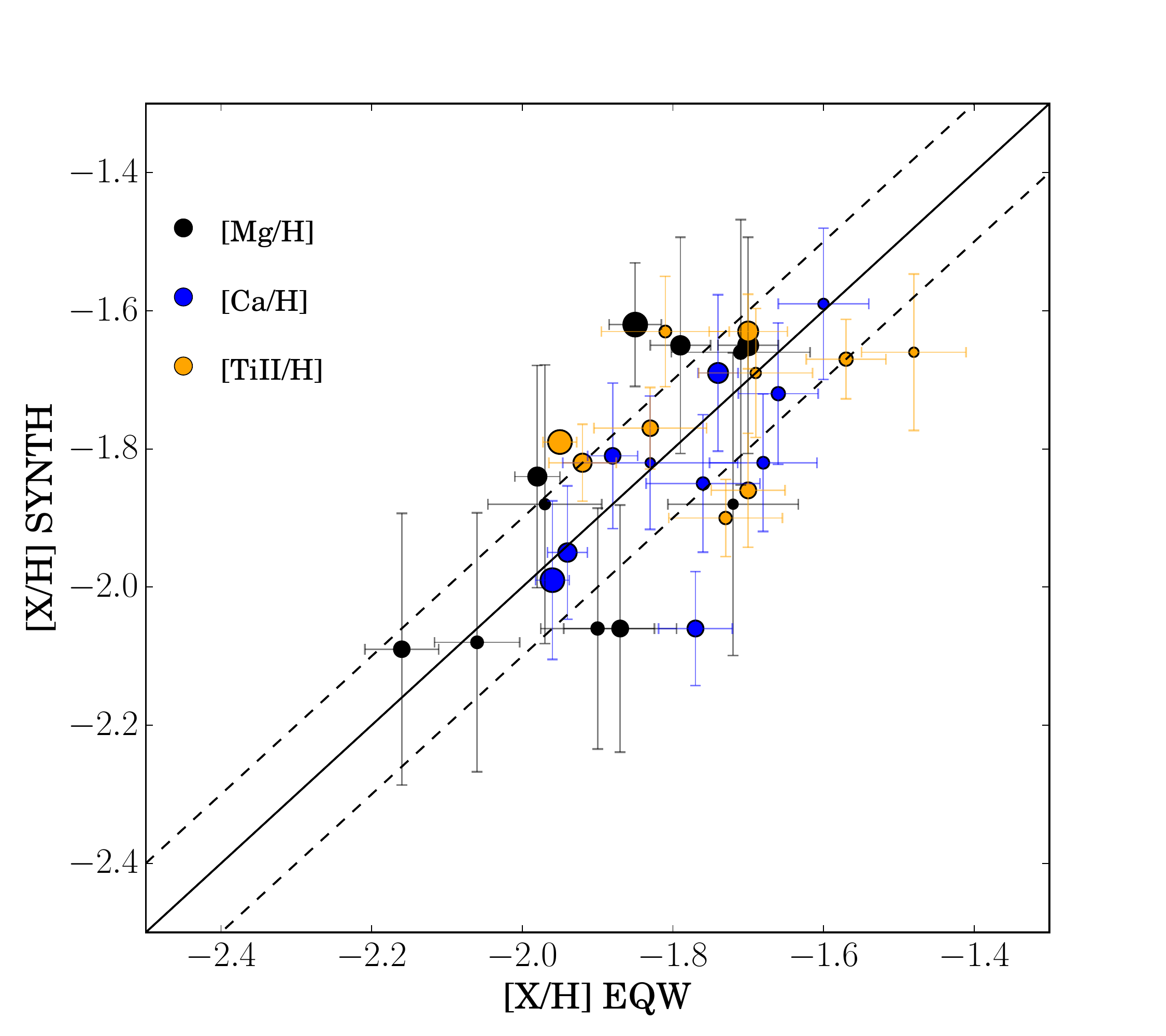}
\caption{Comparison between the abundances of three $\alpha$-elements derived by the EQW  and SYNTH methods. The
solid line traces the line of equality while the dashed lines indicate a
variation by $\pm$ 0.1 dex around this line.}
\label{Comp_XH_EQW_SYNTH_all_El}
\end{figure}

Figure \ref{Comp_XH_EQW_SYNTH_all_El} provides a comparison of the
two techniques for three $\alpha$-elements Mg, Ca, and
\ion{Ti}{II}. These are the elements for which we have the largest number of detections, providing
a reliable comparison. Here, again the abundances resulting from the two techniques are
perfectly consistent within the error bars.

\subsection{Comparison with literature}

\label{comp_giraffe}
\begin{figure}
   \centering
        \includegraphics[width=\hsize]{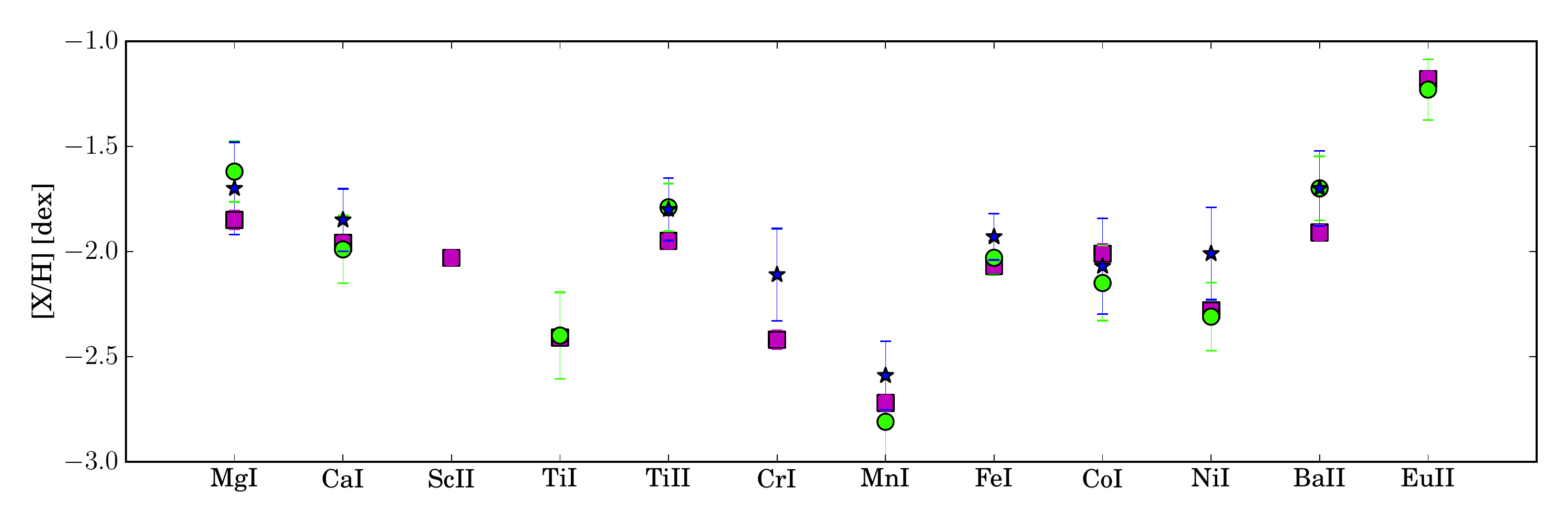}   
      \caption{Elemental abundances of the star S08-3. The green circles refer to this study,
      the blue stars come from \cite{Shetrone2001}, and the magenta squares are from the EQW analysis.}\label{S08-3_common}   
   \end{figure}

S08-3 is the brightest star of the GIRAFFE sample. It has also
been observed with HIRES at the Keck I telescope at a resolution of
R$\sim$34000 and analysed by \citet{Shetrone2001}. There are ten
elements in common between the three independent types of analyses,
SYNTH, EQW in this work, and
\citet{Shetrone2001}. Figure~\ref{S08-3_common} presents a
comparison of the results from these analyses.  The median of the difference between the studies is 0.09
dex with a minimum at 0.01 dex and a maximum at 0.30 dex, all well
within the error bars.

In conclusion, the tests we performed confirm that the choice of one or the
other technique to derive the elemental abundances is a source of only minor variation
in the results, and we did not find any bias or deviation that would exceed the
uncertainties on the abundances. Therefore, we confidently combine our UVES/EQW
and GIRAFFE/SYNTH analyses with previously published results in following
discussion.

\section{Results}

\subsection{Metallicity distribution function}

\begin{figure}
   \centering
   \includegraphics[width=\hsize]{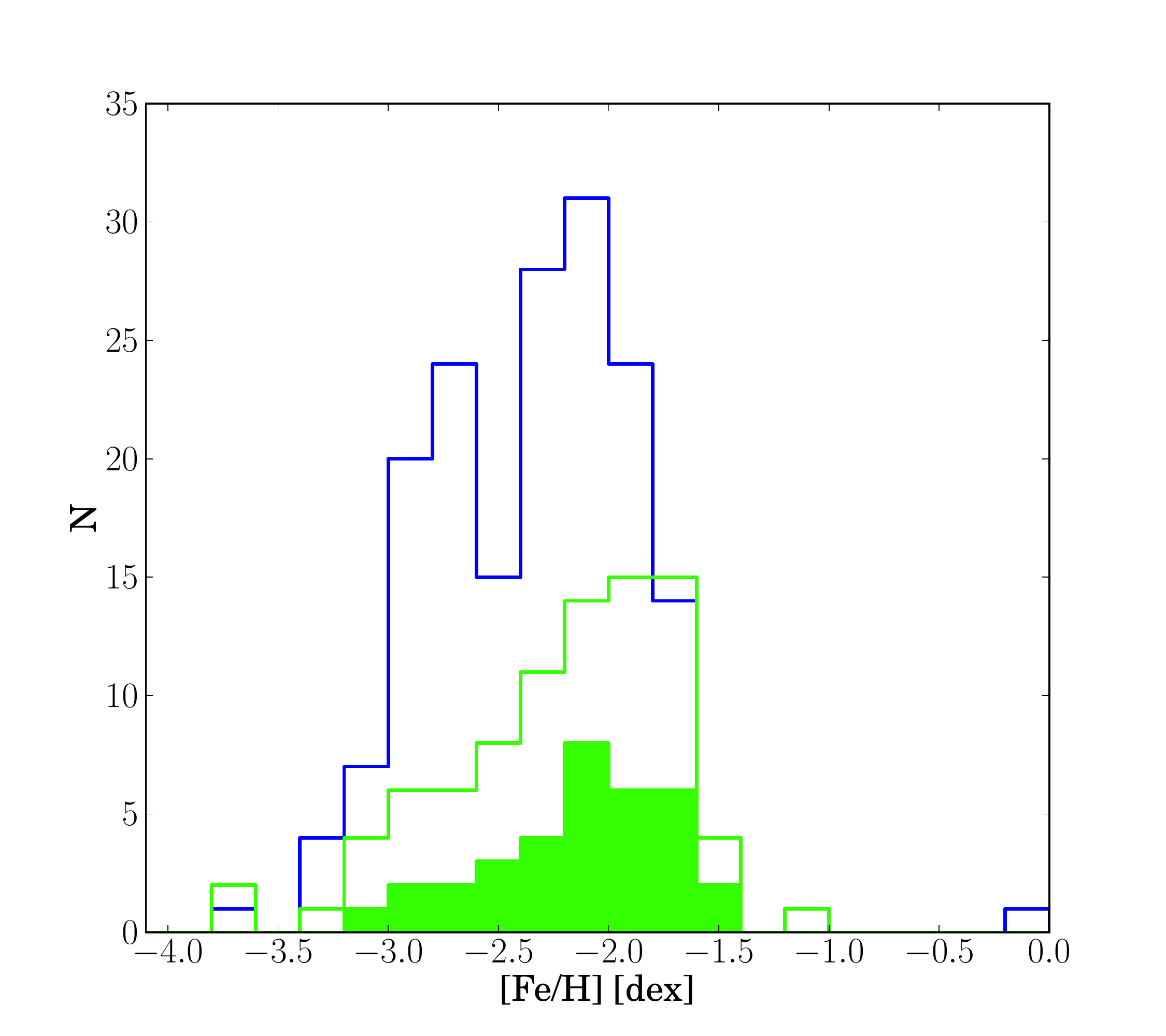}
      \caption{Metallicity distribution function of the Sextans stars. Following the same
      colour code as in Figs. \ref{Star_Location} and \ref{CMD}. The blue solid line
      corresponds to the probable members of the CaT DART sample
      \citep{Battaglia2011}, while the green solid line shows our high-resolution sample. The filled
      green area corresponds to the subsample of our members previously studied in CaT.}
         \label{MDF}
   \end{figure}

The metallicity distribution function of our FLAMES sample stars is presented in
Figure \ref{MDF}. The Sextans members (green solid line) are compared to the
DART medium-resolution Ca triplet survey \citep[blue solid
  line;][]{Battaglia2011}. Our distribution peaks around [Fe/H] = $-1.8$ dex and
is almost completely included in the CaT distribution. Keeping only the stars
with CaT metallicity estimates, we obtain the filled green area peaking at the
same [Fe/H] value as the original CaT distribution.  Our full dataset is on
average slightly more metal-rich than the full CaT DART survey. This is most
likely a consequence of the different spatial distribution of the two samples
(see Figure \ref{Star_Location}) and the presence of a radial gradient in
Sextans \citep{Lee2009, Battaglia2011}. Our field is centrally concentrated
(diameter of 25\arcmin), while the DART CaT survey covers FLAMES fields
distributed from the centre to near the tidal radius of Sextans.
\citet{Revaz2018} have shown that gas tend to concentrate in the central regions
of the dwarf spheroidal galaxies which consequently have the longest star
formation activity. Indeed the analysis of the colour-magnitude diagrams
indicates progressively longer timescales of star formation towards the galaxy
centre \citep{Lee2009}. Therefore, our central FLAMES field is probing the full
chemical evolution of Sextans as illustrated by the fact that we cover a very
large metallicity range, from $-3.2$ to $-1.5$~dex.

Among the stars with $S/N \ge 10$ spectra, two are found at [Fe/H] $\le -3:$
S05-94 and S08-257. Two more have [Fe/H] $\sim -2.90$, namely S08-71 and S11-97, which
makes them eligible, within the uncertainties, for membership of the class of extremely
metal-poor stars. Indeed, S11-97 has been reobserved with UVES at a
resolution of $\sim 40 000$ and is confirmed as an EMP (Lucchesi et al., in
prep).

\subsection{$\alpha$-elements}
\label{alpha}

\begin{figure}
\centering
\includegraphics[width=\hsize]{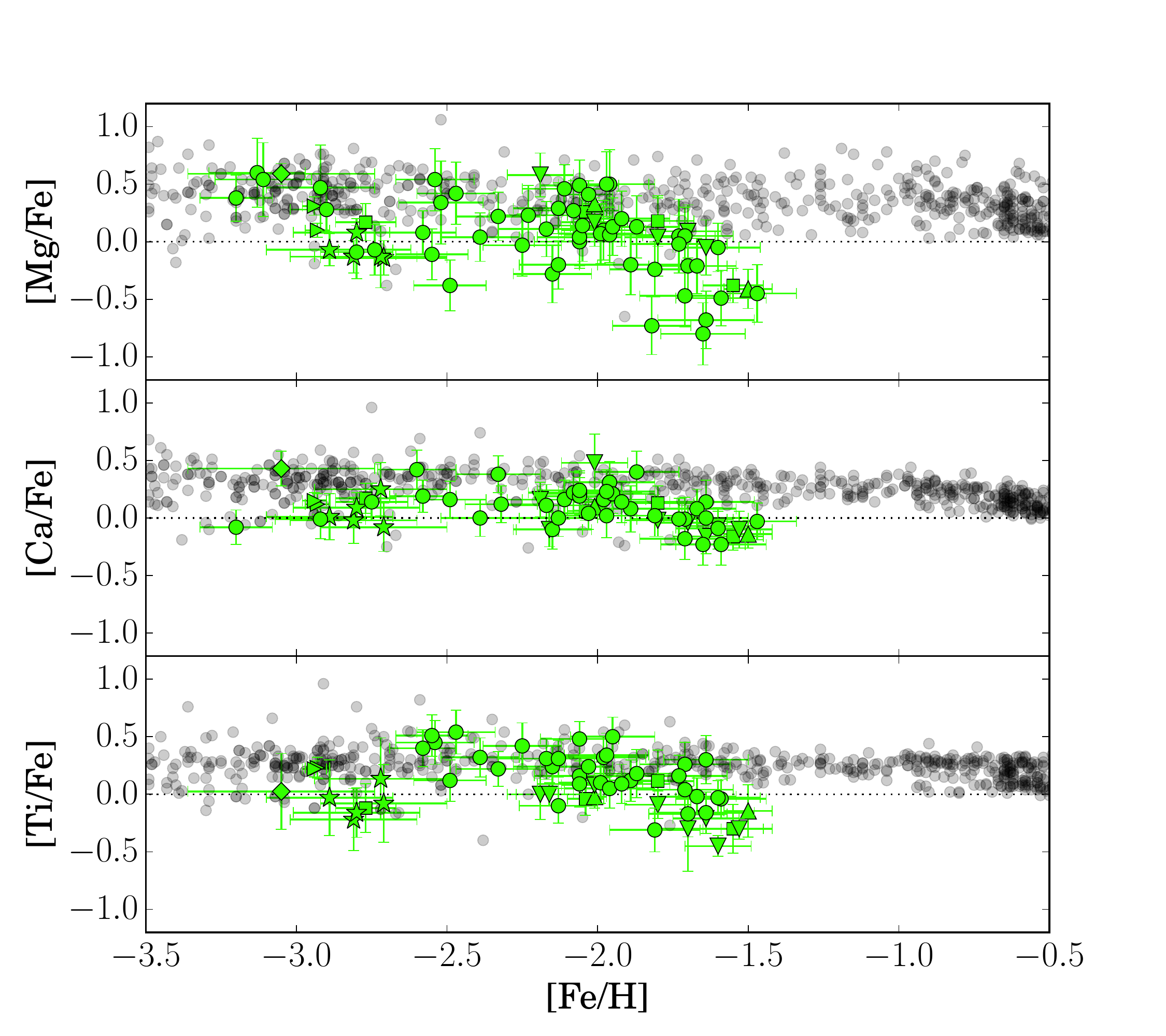} 
\caption{Distribution of [Mg/Fe], [Ca/Fe], and [Ti/Fe] as a function
  of [Fe/H] for the Sextans dSph (in green).
The circles are the new stars
  from our Giraffe sample; the squares display the re-analysed sample
  of \citet{Shetrone2003}; the stellar symbols stand for the dataset of
  \citet{Aoki2009}, the down-pointing triangles for \citet{Kirby2010II}, the right-pointing
  triangles for \citet{Tafelmeyer2010}, and the diamond from
  \citet{Honda2011}. The Milky Way comparison sample shown in grey encompasses 
\citet{Venn2004, Cayrel2004, Francois2007, Gratton2003, Cohen2006, Cohen2008, Cohen2013, Honda2004B, Reddy2006, Yong2013, Lai2007, Spite2005, Aoki2005, Aoki2007, Barklem2005, Ishigaki2013}.}
\label{AlphaFe}
\end{figure}

Three $\alpha$-elements are accessible in our wavelength range:
\ion{Mg}{I} (1 line), \ion{Ca}{I} (13 lines), and Ti at its two
ionization levels, \ion{Ti}{I} (4 lines) and \ion{Ti}{II} (4
lines). The distribution of [$\alpha$/Fe] as a function of [Fe/H] is
presented in Fig.~\ref{AlphaFe} for Sextans and the Milky Way
stars. Only stars with S/N$>$10 spectra are considered.  Our study
significantly increases the existing samples at high resolution with 46 new
stars in [Mg/Fe], 37 in [Ca/Fe], and 34 in [Ti/Fe].

The comparison samples taken from the literature come from works
conducted at high spectroscopic resolution, at the exception of
\citet{Kirby2010II} with R$\sim$7000. In order to keep the level of
accuracy in abundance ratios as homogeneous as possible between the
different references, we restricted the sample of \citet{Kirby2010II}  to
the eight stars with error on [Fe/H] $\leq 0.12$ dex. The star S15-19 was analysed by both \citet{Aoki2009} and \citet{Honda2011}. We
consider the abundances of the latter study. We refer the reader to
Lucchesi et al. (in prep) for a full reanalysis of the sample of
\citet{Aoki2009}.

\subsubsection{\ion{Ti}{I} versus \ion{Ti}{II}}
Due to the weakness of the \ion{Ti}{I} lines, we were only able to derive the [\ion{Ti}{I}/Fe] abundance
ratio  in four stars (S05-47, S08-3, S08-6, S08-38), while
[\ion{Ti}{II}/Fe] could be calculated in 32 stars. Still, the result on
[\ion{Ti}{I}/Fe] for S05-47 is extremely uncertain as it relies on a single
line.  The three other stars show lower [\ion{Ti}{I}/Fe] than
[\ion{Ti}{II}/Fe] but the estimates  agree within the error bars.
In the following, [Ti/Fe] is taken as  [\ion{Ti}{II}/Fe] similarly to our previous studies.

\subsubsection{Comparison between $\alpha$-elements and impact of the SNeIa}
Figure \ref{AlphaFe} shows the common trend between the three $\alpha$-elements and
[Fe/H]: a plateau at constant [$\alpha/Fe$] with [Fe/H] followed by a
  decrease down to subsolar values.   The scatter in abundance ratios
  scales with the number of lines of each element usable in the analysis and the
  S/N of the spectra. This is particularly true for [Mg/Fe], for which the
scatter at fixed [Fe/H] and the error bars are the largest.  Regarding the stars
  with subsolar [Mg/Fe] values, three are particularly noticeable, S08-282,
  S08-280, and S05-47, in order of increasing [Fe/H], with [Mg/Fe] $< -0.5$.
While the exact Mg abundance for these stars is not yet secure,  as
  uncertainties are large, the comparison with other stars in the sample at
similar atmospheric parameters confirms their low values: S08-282 (\teff=5044K,
\logg=1.96, 1.61, [Fe/H]=$-1.96$) was compared to S05-84 (5048K, 1.92, 1.62,
$-1.96$; [Mg/Fe] = 0.5), and S05-47 (4654K,1.28, 1.74, $-1.64$) to S08-242
(4657K, 1.44, 1.44, $-1.71$; [Mg/Fe] = +0.05).

The trend seen in Fig.~\ref{AlphaFe} is formed by a plateau at
[$\alpha$/Fe]$\sim 0.4$ ended by the so-called knee, which is the [Fe/H] origin
of the decrease in [$\alpha$/Fe] with increasing metallicity, corresponding to
the time when, after a decrease in the galaxy star formation rate, the
contribution of the SNeIa starts to dominate the chemical composition of the
ISM. This new FLAMES sample offers for the first time a large
number of stars with abundances derived at sufficient accuracy to estimate the
location of this knee in Sextans at [Fe/H] $\sim -2$~dex.

The galaxy star formation history shapes the morphology of the [$\alpha$/Fe]
versus [Fe/H] diagram. The similar position of the knee reveals similar star formation
efficiency in the first few gigayears of the galaxy evolution, while the slope of the
decrease in [$\alpha$/Fe] is determined by the balance between the mass of
metals ejected by the SNeIa on the one hand and by the SNeII on the other
hand. Therefore, past the peak of its star formation rate, a system with
extended star formation history will lead to a decrease in [$\alpha$/Fe] versus
[Fe/H] with a smaller slope than a galaxy that is quenched sharply. As a
consequence, it is quite informative to compare the decreasing [$\alpha$/Fe]
branches of the classical dwarfs.

The { position of the knee in the Carina dSph is still  an open issue,
most probably because of the overlap of stellar populations coming from the
three different star formation episodes. It was tentatively detected between
$-2.7$ and $-2.3$~dex \citep{Lemasle2012, Venn2012}, however not confirmed by
\cite{Norris2017}. The spectroscopic samples of other dSphs, such as those of
Draco and Ursa Minor, are still too small to robustly locate their knees
\citep{CohenHuang2009, CohenHuang2010}. The Fornax and Sculptor dSphs, for which
there exists sufficiently large spectroscopic samples, are therefore the two
systems that are comparable to Sextans \citep{Letarte2010, hill2019}.  The
Sextans and Sculptor dSphs have formed stars for 4 to 6 Gyrs, respectively
\citep{Lee2009,deBoer2011}, while the period of star formation in Fornax is more
extended, beyond 12 Gyr even if at a low final rate \citep{deBoer2012b}. Figure
\ref{AlphaKnee} compares the corresponding [$\alpha$/Fe] versus [Fe/H] patterns,
and provides a detailed view of the knee region.  Formally, the position of the
Sextans knee is slightly below those of Sculptor \citep[$\sim
  -1.8$][]{Tolstoy2009}, and Fornax \citep[$\sim -1.9$][]{Hendricks2014a}.  In
practice these values are essentially identical given the uncertainty on both
[Fe/H] and on the exact position of the knee due to the scatter in [$\alpha$/Fe]
at fixed metallicity. This is an impressive agreement given that these galaxies
span a factor of $\sim$50 in final stellar masses and a range of dynamical
status (Sculptor 2.3 $\times$ 10$^6$ M$_{\star}$ ($\sigma$ = 9.2 km/s), Sextans
0.44$\times$ 10$^6$ M$_{\star}$ ($\sigma$ = 7.9 km/s), Fornax 20$\times$ 10$^6$
M$_{\star}$ ($\sigma$ = 11.7 km/s); \cite{McConnachie2012}).  It provides
evidence that, in the first gigayears, the star formation efficiency (the mass
of newly born stars per unit gas mass) of the three galaxies was very
similar. This period of time corresponds to the initial merging sequence of
small galactic units building the galaxy potential well, which will then be
massive enough (or not) to resist the UV-background heating
\citep{Revaz2018}. In other words, up to the knee, the star formation processes
are essentially dominated by the smaller building blocks. The chemical evolution
past the knee then reflects the in-situ star formation when the mass of the
systems is already in place. This part depends on the degree to which the
galaxies have been impacted (quenched) by the UV background.  As expected from
the length of their star formation histories, the slope of the [$\alpha$/Fe]
versus [Fe/H] branches is steeper for Sextans than for Sculptor and Fornax.

\begin{figure}
\centering
\includegraphics[width=\hsize]{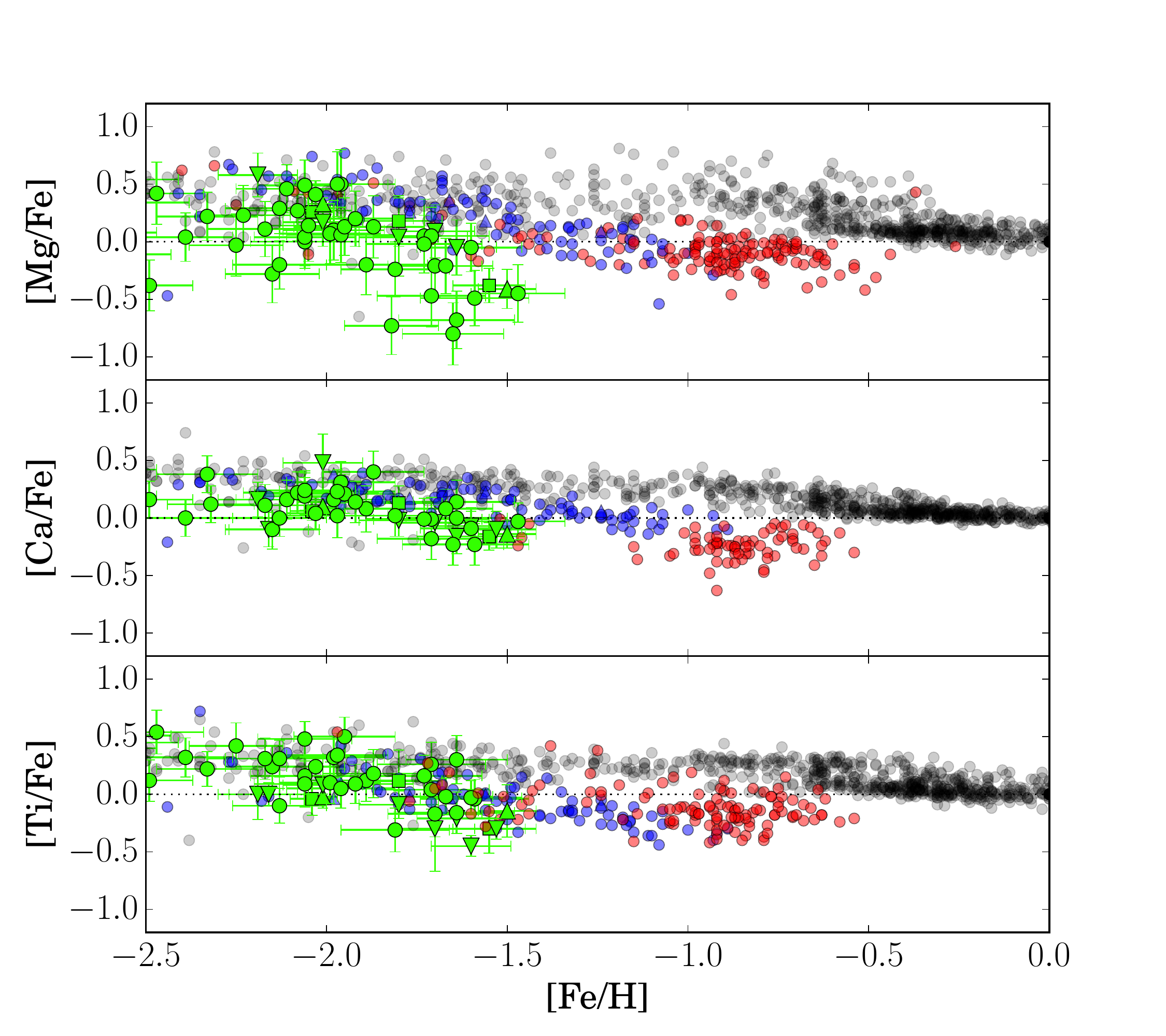} 
\caption{Comparison between the three classical dwarfs, Sextans (green), Sculptor
  (blue), and Fornax (red), and the Milky Way (grey) in the region of the knee in [$\alpha$/Fe].}
\label{AlphaKnee}
\end{figure}

\subsection{Iron-peak elements}

 The number of stars for which the abundances could be determined is smaller
 than for the $\alpha$ elements: 20 stars for Cr, 5 stars for Mn, 2 stars for Co, and
 6 stars for Ni, as a consequence of the weakness of the lines and the large differences in
 the S/Ns of the spectra.

   \begin{figure}
   \includegraphics[width=\hsize]{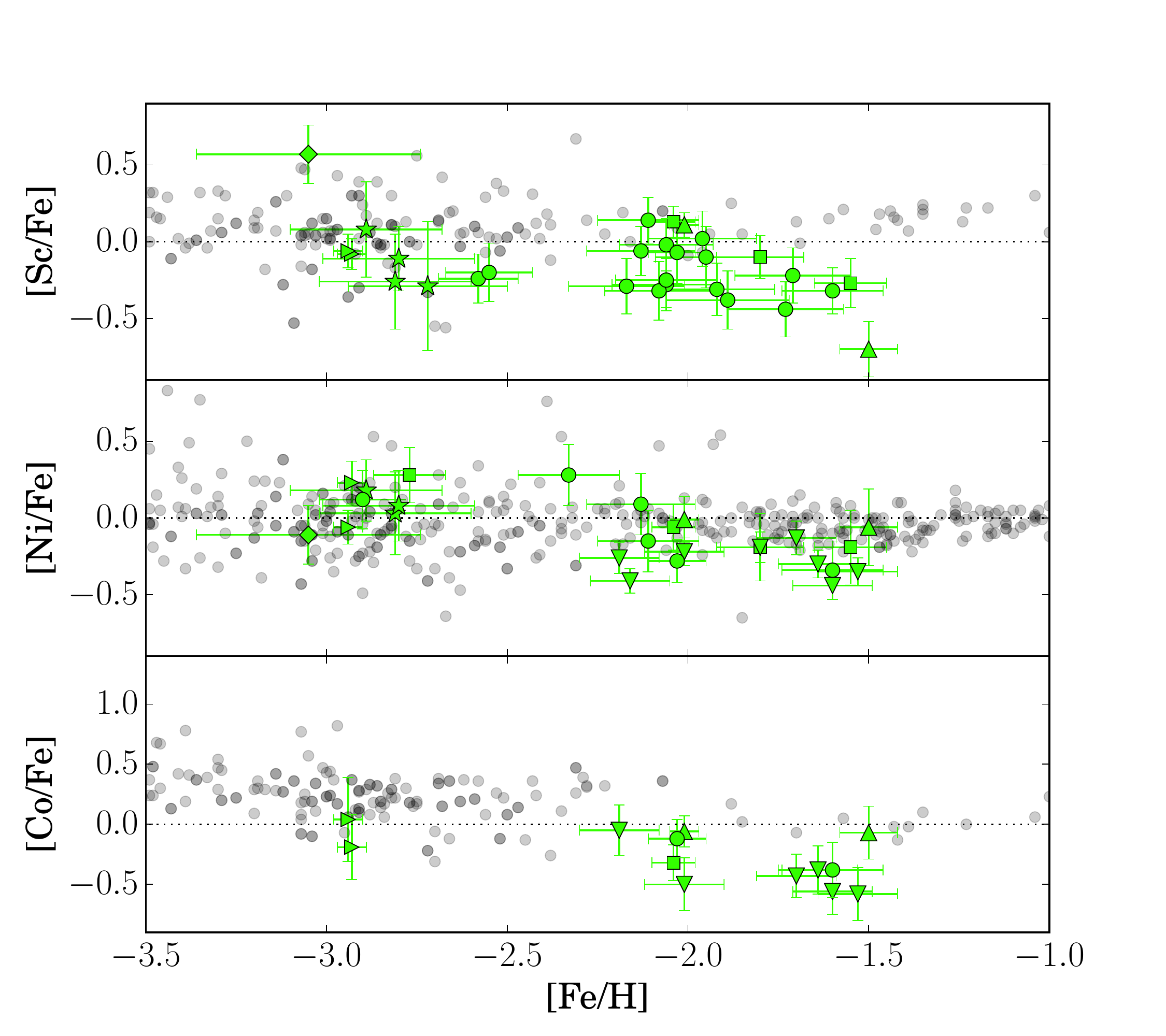}
      \caption{Abundance ratio for the iron-peak elements Sc, Ni, and Sc for the Sextans dSph in green
in comparison with the Milky-Way population in grey. Symbols are as in Fig.\ref{AlphaFe}.}
      \label{ScNiCoFe}
   \end{figure}

Figure \ref{ScNiCoFe} presents the abundance ratios of Sc, Ni, and Co to Fe.
These have been derived from three lines of \ion{Sc}{II}, 6 lines of \ion{Ni}{I},
and one line of \ion{Co}{I}. The production of Sc is dominated by SNeII
\citep{Woosley2002, Battistini2015}, and therefore, as expected, [Sc/Fe] follows the same
trend with [Fe/H] as the $\alpha$-elements. The case of Ni and Co is different
because these elements can also be largely produced by SNeIa
\citep{Timmes2003,Travaglio2005}. The $\sim 2$ Gyr of star formation history
of Sextans, albeit with a sharp drop in SFR beyond the explosion timescale of
SNeIa, makes it an ideal system to witness their nucleosynthesis imprint.
\cite{Kirby2019} conducted an analysis of the Cr, Ni, and Co abundance
trends in dwarf spheroidals in order to constrain the nature of the SNeIa
progenitors. In particular, they compared the theoretical yields of
  Chandrasekhar-mass SNIa and sub-Chandrasekhar-mass models to existing
  observed abundances in dwarf galaxies. Figure \ref{ScNiCoFe} confirms the
decline of [Co/Fe] and [Ni/Fe] past [Fe/H]$\ge -2$.  This implies that the
production of Ni and Co is lower than the yield of Fe in SNeIa. Our conclusions agree with those of
 \cite{Kirby2019}, in that while the observation of [Co/Fe] is
compatible with most of the models, the decrease in [Ni/Fe]  down to
  subsolar values favours the explosion of double degenerate
sub-Chandrasekhar-mass white dwarfs \citep{Shen2018,Bravo2019}.  This nevertheless
requires conformation with NLTE calculations when they become available.

Figure \ref{CrMnFe} shows the two other iron-peak elements that we were able to
measure, \ion{Cr}{I} (3 lines) and \ion{Mn}{I} (4 lines).  \cite{North2012} 
derived the abundance of Mn for the same stars as the present analysis. The results of their study agree with ours within 1$\sigma$ uncertainties. Therefore, for the sake of
homogeneity Fig.~\ref{CrMnFe} displays the results of the synthetic approach of
this work.

Both Cr and Mn closely follow the Milky Way trends. There are no NLTE
calculations available for the range of stellar atmospheric parameters covered
by our sample, either for Cr or for Mn.  Nevertheless,
\citet{BergemannCescutti2010} show that the steady increase of [Cr/Fe]
with metallicity observed for the MW metal-poor stars is an artefact of
neglecting NLTE effects in the line formation of Cr. These NLTE corrections are
positive and move [Cr/Fe] close the solar value over the full range of
metallicities. Given the similar behaviour of the metal-poor population in dwarfs
and the Milky Way, there are good reasons to believe that NLTE is also at the
origin of the low [Cr/Fe] values at [Fe/H] $\le -2$. The fact that [Cr/Fe] is
again solar beyond the position of the knee in Sextans suggests that the
production of chromium is very similar in SNeII and SNeIa.

Departures from LTE have been studied by \cite{Bergemann2008} and
  \cite{Bergemann2019} for slightly hotter stars than our sample and for
different lines from those in our list but in the same metallicity
range. The NLTE abundances of Mn were systematically higher than the
LTE abundances by up to 0.6 dex. This is probably more than needed for
our sample in Sextans to place the sequence of [Mn/Fe] on the solar sequence. if confirmed,
similarly to Cr, the production of Mn should be very similar in SNeII and SNeIa.

   \begin{figure}
   \centering
   \includegraphics[width=\hsize]{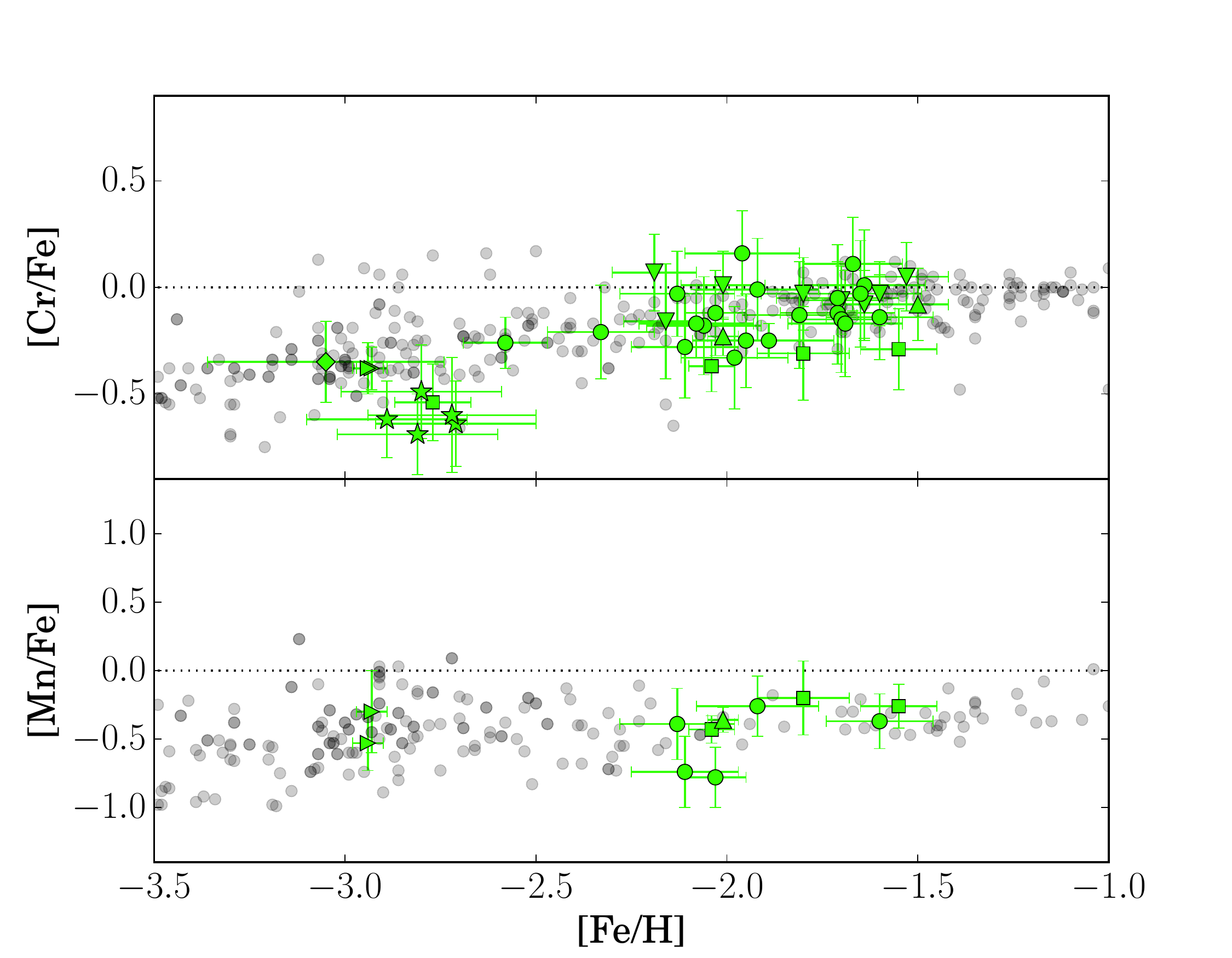}
      \caption{Distribution of [Cr/Fe] (top panel) and [Mn/Fe] (bottom panel) for stars in Sextans and
      the Milky Way. Symbols are as in Fig.~\ref{AlphaFe}.}
         \label{CrMnFe}
   \end{figure}

\subsection{Neutron-capture elements}

We were able to analyse two neutron-capture process elements: \ion{Ba}{II} (2 lines)
measured for 39 stars and \ion{Eu}{II} (1 line) obtained for 7 stars. These
elements are produced in the rapid (r-) and the slow (s-) neutron capture
processes.

   \begin{figure}[!h]
   \centering
   \includegraphics[width=\hsize]{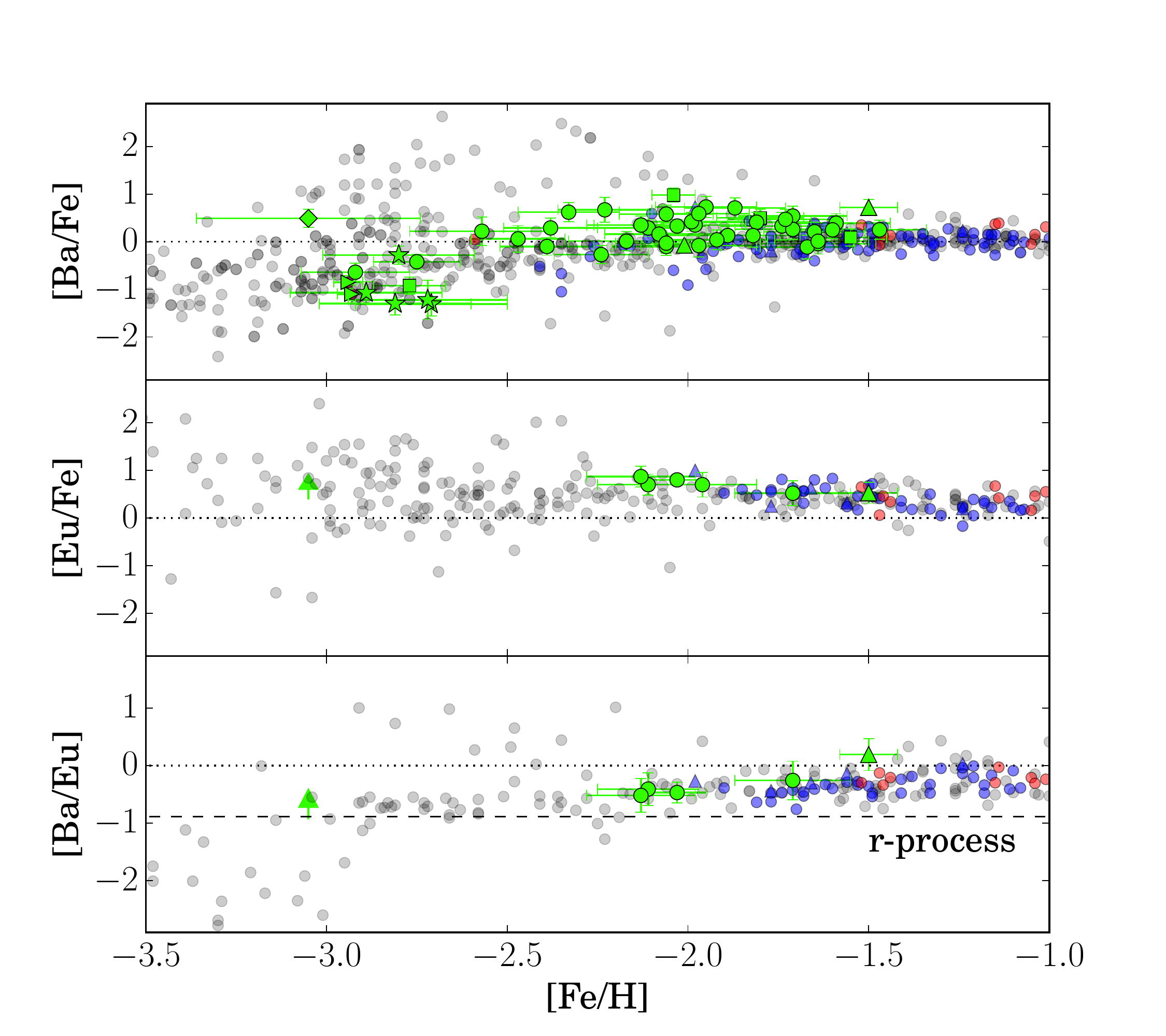}
      \caption{Distributions of [Ba/Fe] (top panel), [Eu/Fe] (middle panel), and
        [Ba/Eu] (bottom panel)  for stars in Sextans (green), Fornax (red),
          Sculptor (blue), and the Milky Way (grey). The upper limit on Eu of
        \cite{Honda2011} for S15-19 at [Fe/H] $\sim -3$ is indicated with an
        arrow.}
         \label{BaEu}
   \end{figure}

Figure \ref{BaEu} presents the evolution of [Ba/Fe], [Eu/Fe], and [Ba/Eu] as a
function of [Fe/H]. The Sextans stars follow the trend of the MW stars in
[Ba/Fe], with a plateau very slightly above the solar level following the $\sim$
1 dex initial scatter between Fe/H]$\sim -3$ and [Fe/H]$\sim -2.5$. We were only
  able to measure europium for a few stars. These sample the solar plateau in
  barium. There is a small but clear decrease of [Eu/Fe] at [Fe/]$\ge -2$,
  indicating that the production of iron progressively overtakes that of Eu, as
  expected from the decrease in star formation rate and the growing impact of
  the explosions of SNeIa.  There is a slight increasing trend of [Ba/Eu] with
  [Fe/H], inversely reflecting the variation of [Eu/Fe] with [Fe/H], with all
  values [Ba/Eu] below [Fe/H] $\sim -1.5$ being subsolar indicating that below
  [Fe/H]$\sim -2$, barium is produced either entirely or mainly by the
  r-process. The low-metallicity asymptotic giant branch stars (AGBs) are
  sufficiently numerous at Fe/H]$\ge -2$ that [Ba/Eu] increases by the addition
    of the s-process elements. Sextans's chemical evolution does not extend as
    far as that of Sculptor; in particular, there is no star in the $-1.5$ to
    $-1$ [Fe/H] range.  However, between [Fe/H]= $-2.3$ and $-1.5$ the two
    galaxies share the same chemical trends, implying similar fractions of SNeIa
    and low-metallicity AGBs \citep{hill2019}. The comparison with Fornax is
    more difficult due to the lack of measurements of europium. Nevertheless, in
    the same low-metallicity range, Fornax does show the same plateau at
    [Ba/Fe]$\sim$0, confirming the universal timescale of the onset and rise of
    the s-process \citep{Lemasle2014}

\section{Summary and conclusions}

We present an analysis of the FLAMES dataset, targeting the central
25\arcmin\ region of the Sextans dSph. This dataset is the
third  part of the ESO large program 171.B-0588(A) obtained by the
Dwarf galaxy Abundances and Radial-velocities Team (DART) together with the
Fornax and Sculptor galaxies \citep{Letarte2010,hill2019}.  A total of 101
stars have been gathered with the HR10, HR13, and H14 gratings of the multi-fibre
spectrograph FLAMES/GIRAFFE. Two fibres were linked to the red arm of the UVES
spectrograph.

Our sample is composed of RGB stars down to V$\sim$20.5 mag, the
level of the horizontal branch in Sextans. Spectra in the
CaT region are also available for some of them \citep{Battaglia2011}, with first estimates of their
metallicity. The rest of the sample was photometrically selected as
corresponding to the position of Sextans in the $V$,$I$ colour-magnitude
diagram. We provide [Fe/H] determination for 81 stars, which cover the wide
metallicity range [Fe/H]=$-$3.2 to $-$1.5 dex. After a thorough investigation
of the random and systematic errors, we deliver accurate abundances of three
$\alpha$-elements (Mg, Ca and Ti), five iron-peak elements (Sc, Cr, Mn, Co, and Ni),
and two neutron-capture elements (Ba and Eu).
Despite the small stellar mass of Sextans, its stellar population
reveals a  rich chemical evolution involving core collapse and
Type Ia supernovae, as well as low metallicity AGBs.

The analysis of the $\alpha$-elements reveals a plateau at
[$\alpha$/Fe]$\sim$0.4 followed by decrease, corresponding to the time when the
contribution of the SNeIa begins to dominate the chemical composition of the
ISM. This new FLAMES sample offers for the first time a  large
number of stars with abundances derived at sufficient accuracy to estimate the
location of the knee in Sextans at [Fe/H]$\sim -2$ dex, very close
to both the Sculptor and Fornax dSphs, despite their
very different masses and star formation histories. This provides evidence that,
in the first gigayears, the star formation of the three galaxies followed similar
processes, such as the accretion of smaller building-blocks before the period
of reionisation.

[Sc/Fe] follows the same trend with [Fe/H] as the $\alpha$-elements,
confirming that its production is dominated by SNeII. Ni and Co can be
produced by SNeII and SNeIa.  We confirm the decline of [Co/Fe] and
[Ni/Fe] past [Fe/H]$\ge -2$, implying that the production of Ni and
Co in SNeIa is lower than that of Fe.  While the observation of
[Co/Fe] is compatible with most of the models of SNeIa, the decrease
in [Ni/Fe] favours the explosion of double degenerate
sub-Chandrasekhar-mass white dwarfs. The fact that [Cr/Fe] is solar
and [Mn/Fe] is flat beyond the position of the knee in Sextans
suggests that the production of chromium and manganese is very similar
in SNeII and SNeIa.

The stellar population of Sextans follows the same trend as the Milky Way stars
in [Ba/Fe], with a plateau at solar level following the initial scatter below
[Fe/H]$\sim -3$. The trend of [Ba/Eu] with [Fe/H] indicates that below [Fe/H]
$\sim -2$, barium is produced either entirely or at least principally by the rapid neutron
capture channel. The low- metallicity AGBs are sufficiently numerous at Fe/H]
  $\sim -2$  that [Ba/Eu] increases by the addition of the s-process elements. In the
  metallicity region covered by the three galaxies, Sculptor, Fornax, and Sextans
  share the same trends, suggesting similar fractions of SNeIa and low-metallicity
  AGBs in their early evolution.

\begin{acknowledgements} The authors are indebted to the International Space Science Institute (ISSI),
Bern, Switzerland, for supporting and funding the international teams "First
stars in dwarf galaxies" and "Pristine''.  RT, PJ, PN, and CL also thank the Swiss
National Science Foundation for its support.
\end{acknowledgements}

\bibliographystyle{aa} 
\bibliography{bibliography} 

\newpage


\begin{landscape}
\begin{longtable}{|l|c|c|c|c|c|c|c|c|l|}
\caption[{Position, photometry and radial velocities of our targets.}]{\label{Photo_param} Position, photometry and radial velocities of our targets. The $V$ and $I$ magnitudes are from ESO/WFI, while the  $J$, $H$ and $K$ magnitudes from UKIRT/WFCAM. The radial velocities were determined using cross-correlation with an input line list thanks to the program DAOSPEC \citep{StetsonPancino2008}. The radial velocities given here are the average between the three values obtained in each grating, therefore we add a tilde  in the case of probable binaries. In the last column we comment on particularities noticed for some stars of our sample : non-member, binary, polluted spectra (by a calibration lamp), carbon star, or incomplete spectra.}\\
\hline
ID & RA & Dec & V & I & J & H & K & v$_{r, mean}$ & Comments \\
  & hms & dms & mag & mag & mag & mag & mag & km/s &   \\
\hline
\endfirsthead
\caption[]{continued.}\\
\hline
ID & RA & Dec & V & I & J & H & K & v$_{r, mean}$ & Comments \\
  & hms & dms & mag & mag & mag & mag & mag & km/s &   \\
\hline
\endhead
\hline
\endfoot
\hline
S05-5   & +10 12 41.6 & $-$01 32 51.6 & 17.59 & 16.32 & 15.22 & 14.60 & 14.48 & 225.7 $\ $ $\pm$ 0.3  & UVES \\
S08-229 & +10 13 10.2 & $-$01 40 50.9 & 18.30 & 17.26 & 16.27 & 15.72 & 15.58 & 221.8 $\ $ $\pm$ 0.4  & UVES \\
S05-10  & +10 12 39.0 & $-$01 29 58.9 & 17.91 & 16.74 & 15.69 & 15.07 & 14.98 & 234.4 $\ $ $\pm$ 0.67 &   \\
S05-47  & +10 12 27.0 & $-$01 30 21.2 & 18.37 & 17.38 & 16.46 & 15.91 & 15.86 & 229.8 $\ $ $\pm$ 1.20 &   \\
S05-60  & +10 13 17.1 & $-$01 26 38.0 & 18.92 & 17.96 & 17.13 & 16.65 & 16.54 & 241.5 $\ $ $\pm$ 1.99 &   \\
S05-67  & +10 12 40.2 & $-$01 29 07.6 & 19.25 & 18.29 & 17.39 & 16.86 & 16.80 & 232.1 $\ $ $\pm$ 1.56 &   \\
S05-70  & +10 12 44.3 & $-$01 33 48.3 & 19.30 & 18.37 & 17.58 & 17.14 & 17.16 & 224.7 $\ $ $\pm$ 2.30 &   Incomplete, not present in HR10 \\
S05-72  & +10 13 09.9 & $-$01 29 34.8 & 19.33 & 18.35 & 17.49 & 16.98 & 17.04 & 233.1 $\ $ $\pm$ 1.87 & Non-member (c), HR13 polluted spectra \\
\end{longtable}
\end{landscape}


\begin{longtable}{l c c c c c r c c c c c }
\caption[{Colour temperatures, bolometric corrections, bolometric magnitudes, \ion{Ca}{II} triplet [Fe/H], signal-to-noise ratios of our targets.}]{\label{Atmospheric_param_from_photometry} The first columns are dedicated to the parameters derived from photometry for the probable members of our sample: the four colour temperatures, the bolometric correction, and bolometric magnitude. [Fe/H]$_{\rm{CaT}}$ is the initial metallicity estimate, either from the  \ion{Ca}{II} triplet calibration if the star was previously observed at medium resolution or taken as the peak metallicity of Sextans, $-1.9$ dex \citet{Battaglia2011}. The last columns provide the S/Ns of the spectra in each of the HR10, HR13, and HR14 grisms, and their average.}\\
ID & T$_{(V-I)}$ & T$_{(V-J)}$ & T$_{(V-H)}$ & T$_{(V-K)}$ & $BC_V$ & M$_{bol}$ & [Fe/H]$_{\rm{CaT}}$ & S/N$_{HR10}$ & S/N$_{HR13}$ & S/N$_{HR14}$ & S/N$_{av}$ \\
  & K & K & K & K & mag & mag & dex &  &  &  &  \\
\hline
\endfirsthead
\caption[]{continued.}\\
\hline
ID & T$_{(V-I)}$ & T$_{(V-J)}$ & T$_{(V-H)}$ & T$_{(V-K)}$ & $BC_V$ & M$_{bol}$ & [Fe/H]$_{\rm{CaT}}$ & S/N$_{HR10}$ & S/N$_{HR13}$ & S/N$_{HR14}$ & S/N$_{av}$ \\
\hline
\endhead
\endfoot
  S05-5 & 4236 & 4247 & 4212 & 4233 &$-$0.66  & $-2.99$ & $-2.31$  $\pm$0.08      &  39  & 40 & 32 & 37\\
S08-229 & 4553 & 4547 & 4515 & 4502 &$-$0.48   & $-2.10$ & $-1.95$ $\pm$0.08      &  18 & 23 & 21 &  21\\
       &       &      &      &      &      &       &               &    &    &    &\\
S05-10 & 4337 & 4350 & 4315 & 4366 & $-$0.57 & $-$2.57 & $-$1.88$\pm$0.08 & 75 & 64 & 75 & 71 \\
S05-47 & 4628 & 4667 & 4626 & 4694 & $-$0.41 & $-$1.96 & $-$1.75$\pm$0.15 & 31 & 25 & 33 & 30 \\
S05-60 & 4711 & 4832 & 4811 & 4830 & $-$0.37 & $-$1.36 & $-$2.24$\pm$0.15 & 22 & 14 & 21 & 19 \\
S05-67 & 4716 & 4764 & 4700 & 4764 & $-$0.39 & $-$1.06 & $-$2.36$\pm$0.30 & 22 & 15 & 31 & 23 \\
S05-70 & 4763 & 4907 & 4934 & 5069 & $-$0.32 & $-$0.94 & \textit{$-$1.90} & 0 & 16 & 25 & 20 \\
\end{longtable}

\begin{longtable}{|l|c|c|c|c|c|c|c|}
\caption[{Radial velocities measured by DAOSPEC in each setup and in average.}]{\label{vrad_HR} The radial velocities measured by DAOSPEC in each of the HR10, HR13, HR14 setups, and their average. Probable binaries are indicated by a tilde symbol. We also provide the number of lines used during the cross-correlation procedure.}\\
\hline
ID & v$_{r, HR10}$ & N$_{HR10}$ & v$_{r, HR13}$ & N$_{HR13}$ & v$_{r, HR14}$ & N$_{HR14}$ & v$_{r, mean}$ \\
  & km/s &   & km/s &   & km/s &   & km/s \\
\hline
\endfirsthead
\caption[]{continued.}\\
\hline
ID & v$_{r, HR10}$ & N$_{HR10}$ & v$_{r, HR13}$ & N$_{HR13}$ & v$_{r, HR14}$ & N$_{HR14}$ & v$_{r, mean}$ \\
  & km/s &   & km/s &   & km/s &   & km/s \\
\hline
\endhead
\hline
\endfoot
\hline
S05-10 & 234.55 $\ $ $\pm$ 0.69 & 37 & 234.58 $\ $ $\pm$ 1.41 & 44 & 234.08 $\ $ $\pm$ 1.25 & 30 & 234.40 $\ $ $\pm$ 0.67 \\
S05-47 & 228.90 $\ $ $\pm$ 0.98 & 37 & 229.94 $\ $ $\pm$ 2.28 & 36 & 230.65 $\ $ $\pm$ 2.58 & 28 & 229.83 $\ $ $\pm$ 1.20 \\
S05-60 & 240.23 $\ $ $\pm$ 1.72 & 34 & 241.68 $\ $ $\pm$ 3.57 & 37 & 242.76 $\ $ $\pm$ 4.45 & 27 & 241.55 $\ $ $\pm$ 1.99 \\
S05-67 & 232.87 $\ $ $\pm$ 2.01 & 33 & 232.37 $\ $ $\pm$ 2.09 & 34 & 231.13 $\ $ $\pm$ 3.68 & 29 & 232.12 $\ $ $\pm$ 1.56 \\
S05-70 & 0.00 $\ $ $\pm$ 0.00 & 0 & 223.26 $\ $ $\pm$ 2.33 & 35 & 226.26 $\ $ $\pm$ 3.96 & 19 & 224.76 $\ $ $\pm$ 2.30 \\
S05-72 & 232.26 $\ $ $\pm$ 2.47 & 36 & 231.43 $\ $ $\pm$ 4.06 & 35 & 235.76 $\ $ $\pm$ 3.00 & 20 & 233.15 $\ $ $\pm$ 1.87 \\
S05-78 & 0.00 $\ $ $\pm$ 0.00 & 0 & 220.12 $\ $ $\pm$ 3.46 & 22 & 223.28 $\ $ $\pm$ 4.81 & 22 & 221.70 $\ $ $\pm$ 2.96 \\
S05-84 & 224.73 $\ $ $\pm$ 3.35 & 35 & 227.27 $\ $ $\pm$ 3.97 & 21 & 225.75 $\ $ $\pm$ 4.31 & 23 & 225.92 $\ $ $\pm$ 2.25 \\
\end{longtable}


%

\tabcolsep 4pt
\scriptsize
\begin{longtable}{lrrr|rr}
\caption{\label{tab:uves_eqw} Equivalent widths of the stars observed with UVES. The asterisk indicates lines with equivalent widths $> 180\AA$ which were not considered in the analysis.}\\
 $\lambda$&El & kiex & $\log{(gf)}$  & US05-5       &US08$-$229     \\
\hline
\endfirsthead
$\lambda$&El & kiex & $\log{(gf)}$   & US05-5       &US08$-$229     \\
\hline
\endhead
4736.77 & \ion{Fe}{I} & 3.21 & $-$0.75&       -      &      -      \\ 
4871.32 & \ion{Fe}{I} & 2.87 & $-$0.36&       -      &      -      \\ 
4872.14 & \ion{Fe}{I} & 2.88 & $-$0.57&       -      &      -      \\ 
4891.49 & \ion{Fe}{I} & 2.85 & $-$0.11&       -      &      -      \\ 
4903.31 & \ion{Fe}{I} & 2.88 & $-$0.93&       -      &      -      \\ 
4918.99 & \ion{Fe}{I} & 2.87 & $-$0.34&       -      &      -      \\ 
4920.50 & \ion{Fe}{I} & 2.83 &  0.07&       -      &      -      \\ 
4938.81 & \ion{Fe}{I} & 2.87 & $-$1.08&       -      &      -      \\
\end{longtable}                 



\begin{table*}
\caption[{Stellar parameters: effective temperatures, surface gravities, micro-turbulence velocities and metallicities.}]{The four stellar parameters for the probables members of our sample : the effective temperature, the surface gravity, the micro-turbulenc velocity and the metallicity. }
\label{Stellar_param}
\centering
\begin{tabular}[t]{|l|c|c|c|c|}
\hline
ID & T$_{eff}$ & $\log{g}$ & v$_{turb}$ & [Fe/H] \\
  & K & cm/s$^2$ & km/s & dex \\
\hline
S05-5  & 4337 & 0.77 & 1.8  & $-$2.07 \\
S08-229& 4625 & 1.23 & 1.23  & $-$1.50 \\

 &  &  &  &  \\
S05-10 & 4342 & 0.91 & 1.82 & $-$2.11$\pm$0.14 \\
S05-47 & 4654 & 1.28 & 1.74 & $-$1.64$\pm$0.16 \\
S05-60 & 4796 & 1.57 & 1.69 & $-$1.98$\pm$0.14 \\
\hline
\end{tabular}
\begin{tabular}[t]{|l|c|c|c|c|}
\hline
ID & T$_{eff}$ & $\log{g}$ & v$_{turb}$ & [Fe/H] \\
  & K & cm/s$^2$ & km/s & dex \\
\hline
S08-222 & 5009 & 2.17 & 1.57 & $-$1.78$\pm$0.14 \\
S08-225 & 4950 & 2.15 & 1.57 & $-$2.73$\pm$0.27 \\
S08-239 & 4546 & 1.34 & 1.73 & $-$1.89$\pm$0.17 \\
S08-241 & 4689 & 1.44 & 1.71 & $-$2.17$\pm$0.16 \\
S08-242 & 4657 & 1.44 & 1.71 & $-$1.71$\pm$0.16 \\
S08-245 & 4803 & 1.56 & 1.69 & $-$2.58$\pm$0.11 \\
\hline
\end{tabular}
\end{table*}

\clearpage


\begin{table*}
\caption{Abundances for S05-5, S08-229 observed with UVES \label{tab:UVES_abund}
}
\small
  \begin{tabular}{rcrrrrrrrrrrrrrr}
    ID  &         &   [\ion{Fe}{I}/H] & [\ion{Fe}{II}/H] & [Mg/Fe]& [Ca/Fe]  & [Sc/Fe]& [\ion{Ti}{I}/Fe]& [\ion{Ti}{II}/Fe] & [Cr/Fe]  & [Mn/Fe] & [Co/Fe] & [Ni/Fe] & [Ba/Fe]  & [Eu/Fe] \\
    \hline

   S05-5 &         & $-$2.01           & $-$1.90          &   0.33 &  0.10   &   0.11 &  $-$0.13        &  0.09             &  $-$0.23 & $-$0.36 & $-$0.06 & $-$0.01 &  $-$0.07 &    \\
         & Err     &  0.04             &  0.05            &   0.13 &  0.06   &   0.08 &  0.07           &  0.07             &  0.09    &  0.09   &   0.13  &  0.15   &   0.08   &     \\
\hline
S08-229 &          & $-$1.50           & $-$1.38         &  $-$0.41 & $-$0.14 & $-$0.70& $-$0.21         & $-$0.08           & $-$0.08 &           & $-$0.07 & $-$ 0.06 &   0.72 &    0.53 \\
        & Err      &  0.08             &  0.08           &   0.17   &  0.12   &  0.18  &  0.17           &  0.15             &  0.17   &           &  0.22   &  0.25    &   0.17 &    0.22 \\
  \end{tabular}
\end{table*}

\begin{table*}[ht!]
 \caption[]{Errors due to uncertainties in atmospheric parameters for S05-5, S08-229 observed with UVES.  They are 
related to [Fe/H] for iron and to [X/Fe] for the other elements.}
\label{tab:atmoerr}}
{\scriptsize
\begin{center}
\begin{tabular}{l|rrr|rrr|}  
&\multicolumn{3}{c}{S05-5}&\multicolumn{3}{c}{S08-229}\\
\hline
EL &  $\Delta$\teff &  $\Delta$\logg  &  $\Delta$\vt   &  $\Delta$\teff &  $\Delta$\logg  &  $\Delta$\vt \\
& 150K & 0.3  &0.2km/s& 150K & 0.3  &0.2km/s\\
\hline
\ion{Ba}{II} & $-$0.16&   0.12&$-$0.07&$-$0.17&   0.09&$-$0.18\\
\ion{Ca}{I}  & $-$0.01&   0.00&   0.04&$-$0.08&$-$0.02&   0.06\\
\ion{Co}{I}  &    0.06&   0.01&   0.05&   0.01&   0.01&   0.06\\
\ion{Cr}{I}  &    0.13&   0.04&$-$0.09&   0.07&$-$0.04&$-$0.13\\
\ion{Cu}{I}  &    0.05&   0.01&   0.05&       &       &       \\
\ion{Eu}{II} &        &       &       &$-$0.24&   0.11&   0.07\\
\ion{Fe}{I}  &    0.20&   0.01&$-$0.07&   0.23&   0.02&$-$0.09\\
\ion{Fe}{II} & $-$0.09&   0.11&$-$0.04&$-$0.10&   0.13&$-$0.05\\
\ion{Mg}{I}  & $-$0.08&$-$0.03&   0.00&$-$0.10&$-$0.04&   0.06\\
\ion{Mn}{I}  &    0.09&   0.03&   0.04&$-$0.02&$-$0.02&   0.07\\
\ion{Na}{I}  &    0.09&   0.03&$-$0.05&       &       &       \\
\ion{Nd}{II} & $-$0.16&   0.10&   0.06&$-$0.20&   0.10&   0.05\\
\ion{Ni}{I}  &    0.01&   0.03&$-$0.14&   0.01&   0.00&$-$0.14\\
\ion{O}{I}   & $-$0.15&   0.12&   0.06&       &       &       \\
\ion{Si}{I}  &        &       &       &$-$0.19&   0.01&   0.09\\
\ion{Ti}{I}  &    0.15&   0.03&$-$0.01&   0.11&   0.02&   0.03\\
\ion{Ti}{II} & $-$0.22&   0.09&$-$0.00&$-$0.27&   0.09&   0.04\\
\ion{Y}{II}  & $-$0.20&   0.09&   0.02&$-$0.22&   0.09&$-$0.03\\
\ion{Zn}{I} & $-$0.25&   0.02&   0.02&$-$0.26&   0.04&   0.03\\
\end{tabular}                                                                                                            
\end{center}
\end{table*}

\begin{table*}
\caption[{Line list of the studied elements.}]{The complete line list used to derive all the atomic abundances : wavelength, species, excitation potential, oscillator strength and C6 constant. The hyperfine components are indicated in italic and their corresponding equivalent lines are followed by \textit{(equi)}.}
\label{Linelist}
\centering
\begin{tabular}{|l|c|c|c|c||c|c|c|c|c|}
\hline
$\lambda$ & El & $\chi_{exc}$ & $\log{(gf)}$ & $C_6$ & $\lambda$ & El & $\chi_{exc}$ & $\log{(gf)}$ & $C_6$ \\
\AA &   & eV &  &  & \AA &   & eV &  &  \\
\hline
5528.410 & \ion{Mg}{I} & 4.350 & $-$0.360 & 1.63e-30 & 5336.770 & \ion{Ti}{II} & 1.580 & $-$1.700 & 0.00e+00 \\
5349.470 & \ion{Ca}{I} & 2.710 & $-$0.310 & 0.00e+00 & 5381.010 & \ion{Ti}{II} & 1.570 & $-$1.780 & 0.00e+00 \\
5581.970 & \ion{Ca}{I} & 2.520 & $-$0.710 & 0.00e+00 & 5418.770 & \ion{Ti}{II} & 1.580 & $-$2.110 & 0.00e+00 \\
5588.750 & \ion{Ca}{I} & 2.520 & 0.210 & 0.00e+00 & 6559.590 & \ion{Ti}{II} & 2.050 & $-$2.020 & 0.00e+00 \\
5590.110 & \ion{Ca}{I} & 2.520 & $-$0.710 & 0.00e+00 & 5345.800 & \ion{Cr}{I} & 1.000 & $-$0.980 & 0.00e+00 \\
5601.280 & \ion{Ca}{I} & 2.520 & $-$0.690 & 0.00e+00 & 5409.800 & \ion{Cr}{I} & 1.030 & $-$0.720 & 3.92e-32 \\
\hline
\end{tabular}
\end{table*}

\begin{table*}
\caption*{Table~\ref{Linelist}: continued.}
\label{Linelistb}
\centering
\begin{tabular}{|l|c|c|c|c||c|c|c|c|c|}
\hline
$\lambda$ & El & $\chi_{exc}$ & $\log{(gf)}$ & $C_6$ & $\lambda$ & El & $\chi_{exc}$ & $\log{(gf)}$ & $C_6$ \\
\AA &   & eV &  &  & \AA &   & eV &  &  \\
\hline
5516.699 & \ion{\emph{Mn}}{I} & 2.18 & $-$3.273 & 6.36e-32 & 6518.370 & \ion{Fe}{I} & 2.830 & $-$2.460 & 4.14e-32 \\
5516.709 & \ion{\emph{Mn}}{I} & 2.18 & $-$2.905 & 6.36e-32 & 6581.220 & \ion{Fe}{I} & 1.480 & $-$4.680 & 2.06e-32 \\
5516.718 & \ion{\emph{Mn}}{I} & 2.18 & $-$2.905 & 6.36e-32 & 6609.120 & \ion{Fe}{I} & 2.560 & $-$2.660 & 4.11e-32 \\
5516.728 & \ion{\emph{Mn}}{I} & 2.18 & $-$4.482 & 6.36e-32 & 5483.340 & \ion{Co}{I} (equi) & 1.710 & $-$1.490 & 2.89e-32 \\
5516.743 & \ion{\emph{Mn}}{I} & 2.18 & $-$2.773 & 6.36e-32 & 5483.309 & \ion{\emph{Co}}{I} & 1.711 & $-$2.120 & 2.89e-32 \\
5516.757 & \ion{\emph{Mn}}{I} & 2.18 & $-$2.773 & 6.36e-32 & 5483.334 & \ion{\emph{Co}}{I} & 1.711 & $-$2.261 & 2.89e-32 \\
\hline
\end{tabular}
\end{table*}

\begin{table*}
\caption*{Table~\ref{Linelist} continued.}
\label{Linelistc}
\centering
\begin{tabular}{|l|c|c|c|c|}
\hline
$\lambda$ & El & $\chi_{exc}$ & $\log{(gf)}$ & $C_6$ \\
\AA &   & eV &  &  \\
\hline
6496.896 & \ion{\emph{Ba}}{II} & 0.604 & $-$1.919 & 0.00e+00 \\
6496.900 & \ion{\emph{Ba}}{II} & 0.604 & $-$1.482 & 0.00e+00 \\
6496.906 & \ion{\emph{Ba}}{II} & 0.604 & $-$2.531 & 0.00e+00 \\
6496.908 & \ion{\emph{Ba}}{II} & 0.604 & $-$2.133 & 0.00e+00 \\
6496.888 & \ion{\emph{Ba}}{II} & 0.604 & $-$2.832 & 0.00e+00 \\
6496.912 & \ion{\emph{Ba}}{II} & 0.604 & $-$2.133 & 0.00e+00 \\
\hline
\end{tabular}
\end{table*}

{\small
\begin{longtable}{|l|l|l|l|l|}
\caption[{Abundances for Fe, Mg, Ca and Sc.}]{\label{Abundances_A} Atomic abundances for Fe, Mg, Ca and Sc. The  solar abundances are from \citet{GrevesseSauval1998}. The number of lines used to determined each abundance is indicated in brackets. The asterisks indicate stars with spectra at signal-to-noise $\leq$ 10.}\\
\hline
ID & [Fe/H] $\pm$ err$_{sys}$ $\pm$ err$_{rand}$ (N) & [Mg/Fe] $\pm$ err$_{sys}$ $\pm$ err$_{rand}$ & [Ca/Fe] $\pm$ err$_{sys}$ $\pm$ err$_{rand}$ (N) & [Sc/Fe] $\pm$ err$_{sys}$ $\pm$ err$_{rand}$ (N) \\
\hline
$\log{\epsilon(X)_{\odot}}$ & 7.50 & 7.58 & 6.36 & 3.17 \\
\hline
\endfirsthead
\caption[]{continued.}\\
\hline
ID & [Fe/H] $\pm$ err$_{sys}$ $\pm$ err$_{rand}$ (N) & [Mg/Fe] $\pm$ err$_{sys}$ $\pm$ err$_{rand}$ & [Ca/Fe] $\pm$ err$_{sys}$ $\pm$ err$_{rand}$ (N) & [Sc/Fe] $\pm$ err$_{sys}$ $\pm$ err$_{rand}$ (N) \\
\hline
$\log{\epsilon(X)_{\odot}}$ & 7.50 & 7.58 & 6.36 & 3.17 \\
\hline
\endhead
\hline
\endfoot
\hline
S05-10 & $-$2.11 $\ \ $ $\pm$ 0.14  $\pm$ 0.01  (50) & +0.46 $\ \ $ $\pm$ 0.15  $\pm$ 0.04 & +0.16 $\ \ $ $\pm$ 0.17  $\pm$ 0.01  (12) & +0.14 $\ \ $ $\pm$ 0.15  $\pm$ 0.03  (3) \\
S05-47 & $-$1.64 $\ \ $ $\pm$ 0.16  $\pm$ 0.01  (50) & $-$0.68 $\ \ $ $\pm$ 0.17  $\pm$ 0.09 & +0.14 $\ \ $ $\pm$ 0.18  $\pm$ 0.03  (11) &    \\
S05-60 & $-$1.98 $\ \ $ $\pm$ 0.14  $\pm$ 0.03  (50) & +0.10 $\ \ $ $\pm$ 0.15  $\pm$ 0.15 & +0.16 $\ \ $ $\pm$ 0.16  $\pm$ 0.06  (7) &    \\
S05-67 & $-$1.81 $\ \ $ $\pm$ 0.15  $\pm$ 0.02  (50) & $-$0.24 $\ \ $ $\pm$ 0.16  $\pm$ 0.10 & +0.02 $\ \ $ $\pm$ 0.17  $\pm$ 0.04  (11) &    \\
S05-70 & $-$2.24 $\ \ $ $\pm$ 0.12  $\pm$ 0.05  (30) &    &    &    \\
S05-78 & $-$2.57 $\ \ $ $\pm$ 0.10  $\pm$ 0.17  (30) &    &    &    \\
S05-84* & $-$1.96 $\ \ $ $\pm$ 0.12  $\pm$ 0.06  (50) & +0.50 $\ \ $ $\pm$ 0.13  $\pm$ 0.23 & +0.27 $\ \ $ $\pm$ 0.14  $\pm$ 0.15  (5) &    \\
S05-86 & $-$1.71 $\ \ $ $\pm$ 0.14  $\pm$ 0.03  (50) & $-$0.47 $\ \ $ $\pm$ 0.16  $\pm$ 0.16 & $-$0.18 $\ \ $ $\pm$ 0.17  $\pm$ 0.06  (8) &    \\
\end{longtable}}

{\small
\begin{longtable}{|l|l|l|l|l|}
\caption[{Atomic abundances for \ion{Ti}{I}, \ion{Ti}{II}, Cr and Mn.}]{\label{Abundances_B} Atomic abundances for \ion{Ti}{I}, \ion{Ti}{II}, Cr and Mn. The  solar abundances are from \citet{GrevesseSauval1998}. The number of lines used is indicated in brackets.  The asterisks indicate stars with spectra at signal-to-noise $\leq$ 10.}\\
\hline
ID & [\ion{Ti}{I}/Fe] $\pm$ err$_{sys}$ $\pm$ err$_{rand}$ (N) & [\ion{Ti}{II}/Fe] $\pm$ err$_{sys}$ $\pm$ err$_{rand}$ (N) & [Cr/Fe] $\pm$ err$_{sys}$ $\pm$ err$_{rand}$ (N) & [Mn/Fe] $\pm$ err$_{sys}$ $\pm$ err$_{rand}$ (N) \\
\hline
$\log{\epsilon(X)_{\odot}}$ & 5.02 & 5.02 & 5.67 & 5.39 \\
\hline
\endfirsthead
\caption[]{continued.}\\
\hline
ID & [\ion{Ti}{I}/Fe] $\pm$ err$_{sys}$ $\pm$ err$_{rand}$ (N) & [\ion{Ti}{II}/Fe] $\pm$ err$_{sys}$ $\pm$ err$_{rand}$ (N) & [Cr/Fe] $\pm$ err$_{sys}$ $\pm$ err$_{rand}$ (N) & [Mn/Fe] $\pm$ err$_{sys}$ $\pm$ err$_{rand}$ (N) \\
\hline
$\log{\epsilon(X)_{\odot}}$ & 5.02 & 5.02 & 5.67 & 5.39 \\
\hline
\endhead
\hline
\endfoot
\hline
S05-10 &    &    & $-$0.28 $\ \ $ $\pm$ 0.24  $\pm$ 0.03  (2) & $-$0.74 $\ \ $ $\pm$ 0.26  $\pm$ 0.04  (1) \\
S05-47 & +0.11 $\ \ $ $\pm$ 0.21  $\pm$ 0.19  (1) & $-$0.16 $\ \ $ $\pm$ 0.17  $\pm$ 0.06  (4) & +0.01 $\ \ $ $\pm$ 0.25  $\pm$ 0.08  (2) &    \\
S05-60 &    & +0.32 $\ \ $ $\pm$ 0.15  $\pm$ 0.11  (4) & $-$0.33 $\ \ $ $\pm$ 0.19  $\pm$ 0.24  (3) &    \\
S05-67 &    & $-$0.31 $\ \ $ $\pm$ 0.16  $\pm$ 0.10  (2) & $-$0.13 $\ \ $ $\pm$ 0.22  $\pm$ 0.25  (1) &    \\
S05-70 &    &    &    &    \\
S05-78 &    &    &    &    \\
S05-84* &    & +0.33 $\ \ $ $\pm$ 0.13  $\pm$ 0.20  (2) &    &    \\
S05-86 &    & +0.26 $\ \ $ $\pm$ 0.16  $\pm$ 0.11  (2) & $-$0.12 $\ \ $ $\pm$ 0.25  $\pm$ 0.24  (3) &    \\
\end{longtable}}

{\small
\begin{longtable}{|l|c|c|c|c|}
\caption[{Atomic abundances for Co, Ni, Ba and Eu.}]{\label{Abundances_C} Atomic abundances for Co, Ni, Ba and Eu. The solar abundances are from \citet{GrevesseSauval1998}. The number of lines used  is indicated in brackets. The asterisks indicate stars with spectra at signal-to-noise $\leq$ 10.}\\
\hline
ID & [Co/Fe] $\pm$ err$_{sys}$ $\pm$ err$_{rand}$ & [Ni/Fe] $\pm$ err$_{sys}$ $\pm$ err$_{rand}$ (N) & [Ba/Fe] $\pm$ err$_{sys}$ $\pm$ err$_{rand}$ (N) & [Eu/Fe] $\pm$ err$_{sys}$ $\pm$ err$_{rand}$ \\
  & dex & dex & dex & dex \\
\hline
$\log{\epsilon(X)_{\odot}}$ & 4.92 & 6.25 & 2.13 & 0.51 \\
\hline
\endfirsthead
\caption[]{continued.}\\
\hline
ID & [Co/Fe] $\pm$ err$_{sys}$ $\pm$ err$_{rand}$ & [Ni/Fe] $\pm$ err$_{sys}$ $\pm$ err$_{rand}$ (N) & [Ba/Fe] $\pm$ err$_{sys}$ $\pm$ err$_{rand}$ (N) & [Eu/Fe] $\pm$ err$_{sys}$ $\pm$ err$_{rand}$ \\
  & dex & dex & dex & dex \\
\hline
$\log{\epsilon(X)_{\odot}}$ & 4.92 & 6.25 & 2.13 & 0.51 \\
\hline
\endhead
\hline
\endfoot
\hline
S05-10 &    & $-$0.15 $\ \ $ $\pm$ 0.19  $\pm$ 0.05  (2) & +0.29 $\ \ $ $\pm$ 0.18  $\pm$ 0.03  (2) & +0.70 $\ \ $ $\pm$ 0.15  $\pm$ 0.06 \\
S05-47 &    &    & $-$0.04 $\ \ $ $\pm$ 0.20  $\pm$ 0.08  (2) &    \\
S05-60 &    &    & +0.35 $\ \ $ $\pm$ 0.18  $\pm$ 0.12  (2) &    \\
S05-67 &    &    & +0.41 $\ \ $ $\pm$ 0.19  $\pm$ 0.08  (2) &    \\
S05-70 &    &    & $-$0.27 $\ \ $ $\pm$ 0.15  $\pm$ 0.12  (2) &    \\
S05-78 &    &    & +0.22 $\ \ $ $\pm$ 0.15  $\pm$ 0.26  (2) &    \\
S05-84* &    &    &    &    \\
S05-86 &    &    & +0.54 $\ \ $ $\pm$ 0.18  $\pm$ 0.11  (2) &    \\
\end{longtable}}


\scriptsize
\tabcolsep 4pt
\begin{longtable}{lllr|cccccccccccc}
  \caption[EQW]{\label{tab:eqw} Equivalent widths of the GIRAFFE sample. The asterisks indicate lines with equivalent widths $> 180\AA$, which were not considered in the analysis (see Section~\ref{sec:eqw})}.\\
 \hline
    $\lambda$&El & kiex & $\log{(gf)}$  &S8-3         &S8-6         &S5-10        &S8-38        &S8-239       &S8-241&        S8-242       &S7-69        &S8-59        &S5-60        &S8-246        &S8-183            \\
\hline
\endfirsthead
\caption[]{continued.}\\
\hline
$\lambda$&El & kiex & $\log{(gf)}$  &S8-3         &S8-6         &S5-10        &S8-38        &S8-239       &S8-241 &         S8-242       &S7-69        &S8-59        &S5-60        &S8-246        &S8-183             \\
\hline
\endhead
\hline
\endfoot
\hline
5339.93 & \ion{Fe}{I} & 3.27 & $-$0.68& 127 \plm 2.4& 144 \plm 3.2& 109 \plm 3.4& 102 \plm 2.6& 119 \plm 4.5&  82 \plm 5.9&  99 \plm 6.1&  82 \plm 3.4&  88 \plm10.2&  46 \plm 4.8&      -      &  87 \plm 8.2\\
5364.86 & \ion{Fe}{I} & 4.45 &  0.22&  78 \plm 1.9& 105 \plm 3.2&  74 \plm 3.3&  50 \plm 2.8&  84 \plm 3.1&  53 \plm 3.7&  67 \plm 2.8&  65 \plm 5.4&      -      &      -      &      -      &  66 \plm 7.7\\
5367.48 & \ion{Fe}{I} & 4.42 &  0.55&  86 \plm 1.9& 107 \plm 3.2&  77 \plm 3.2&  66 \plm 1.9&  98 \plm 3.2&  52 \plm 2.9&  82 \plm 4.8&  72 \plm 4.9&  78 \plm 6.0&  37 \plm 4.2&  80 \plm 6.4&  90 \plm 4.6\\
5369.96 & \ion{Fe}{I} & 4.37 &  0.54& 108 \plm 8.0& 129 \plm 4.6&  95 \plm 5.1&  86 \plm 7.9&  99 \plm 6.0&  47 \plm 4.3&  96 \plm 4.8&  73 \plm 6.7&  73 \plm 5.9&  42 \plm 4.5&      -      &      -      \\
5371.50 & \ion{Fe}{I} & 0.96 & $-$1.64&   ($\ast$)  &   ($\ast$)  &   ($\ast$)  &   ($\ast$)  &   ($\ast$)  & 176 \plm 6.6&   ($\ast$)  & 174 \plm 5.8&   ($\ast$)  & 152 \plm 7.2&   ($\ast$)  & 174 \plm 7.9\\
5383.37 & \ion{Fe}{I} & 4.31 &  0.50& 110 \plm 1.7& 126 \plm 1.9&  99 \plm 2.6&  83 \plm 2.7& 105 \plm 5.3&  65 \plm 4.0&  92 \plm 6.4&  78 \plm 5.0&      -      &  56 \plm 3.8& 108 \plm 4.9&      -      \\
5389.48 & \ion{Fe}{I} & 4.42 & $-$0.40&  51 \plm 3.7&  65 \plm 2.6&  49 \plm 1.8&      -      &      -      &      -      &  41 \plm 1.7&      -      &      -      &      -      &  62 \plm 3.4&      -      \\
5393.17 & \ion{Fe}{I} & 3.24 & $-$0.92& 131 \plm 2.7& 145 \plm 4.4& 114 \plm 3.0&  99 \plm 2.0& 122 \plm 3.6&  78 \plm 5.1& 117 \plm 6.1&  89 \plm 5.7& 110 \plm 5.9&  65 \plm 4.0& 115 \plm 5.4&  77 \plm 4.6\\
\end{longtable}

\end{document}